\documentclass[pdflatex,sn-mathphys-num]{sn-jnl}%

\usepackage{graphicx}%
\usepackage{multirow}%
\usepackage{amsmath,amssymb,amsfonts}%
\usepackage{amsthm}%
\usepackage{mathrsfs}%
\usepackage[title]{appendix}%
\usepackage{xcolor}%
\usepackage{textcomp}%
\usepackage{manyfoot}%
\usepackage{booktabs}%
\usepackage{algorithm}%
\usepackage{algorithmicx}%
\usepackage{algpseudocode}%
\usepackage{listings}%
\DeclareFontFamily{OT1}{pzc}{}
\DeclareFontShape{OT1}{pzc}{m}{it}%
{<-> s * [1.15] pzcmi7t}{}
\DeclareMathAlphabet{\mathpzc}{OT1}{pzc}{m}{it}

\newcommand{\be}{\begin{equation}}

\newcommand{\ee}{\end{equation}}
\newcommand{\bea}{\begin{eqnarray}}
\newcommand{\eea}{\end{eqnarray}}
\newcommand{\beas}{\begin{eqnarray*}}
\newcommand{\eeas}{\end{eqnarray*}}
\newcommand{\nn}{\nonumber}

\newcommand{\eqn}[1]{Eq.~(\ref{#1})}
\newcommand{\fig}[1]{Fig.~\ref{#1}}

\newcommand{\pslash}{p\hspace{-1.5mm}/}

\newcommand{\Cl}[2]{{\mbox{Cl}}_{#1}\left(#2\right)}

\def\s#1{{\scriptscriptstyle #1}}

\def\1eq#1{Eq.~(\ref{#1})}

\def\2eqs#1#2{Eqs.~(\ref{#1}) and~(\ref{#2})}
\def\3eqs#1#2#3{Eqs.~(\ref{#1}),~(\ref{#2}) and~(\ref{#3})}

\def\fig#1{Fig.\ref{#1}}

\newcommand{\GL}{{\Gamma}_{\!\!{\s{\mathbf{L}}}}} 
              
\begin{document}

\title{
Perturbative analysis of the three gluon vertex in different gauges at one-loop}


\author[1]{\fnm{J.Alejandro} \sur{Alfaro}}\email{jesuss@hotmail.com}
\author*[2,3]{\fnm{L. X.} \sur{Guti\'errez-Guerrero}}\email{lxgutierrez@mctp.mx}
\author[3,4]{\fnm{Luis} \sur{Albino}}\email{luis.albino.fernandez@gmail.com}
\author[3,5]{\fnm{Alfredo} \sur{Raya}}\email{alfredo.raya@umich.mx}
\affil[1]{\orgdiv{Facultad de Ciencias en F\'isica y Matem\'aticas}, \orgname{ Universidad Aut\'onoma de Chiapas
}, \orgaddress{\city{Tuxtla Gutiérrez}, \postcode{29040}, \country{México}}}
\affil[2]{\orgdiv{CONACyT-Mesoamerican Centre for Theoretical Physics}, \orgname{ Universidad Aut\'onoma de Chiapas, 
}, \orgaddress{\city{Tuxtla Gutiérrez}, \postcode{29040}, \country{México}}}
\affil[3]{\orgdiv{Instituto de F\'isica y Matem\'aticas}, \orgname{Universidad
Michoacana de San Nicol\'as de Hidalgo}, \orgaddress{\city{Morelia, Michoac\'an}, \postcode{58040}, \country{M\'exico}}}
\affil[4]{\orgdiv{Departamento Sistemas F\'isicos Qu\'imicos y Naturales}, \orgname{Univ. Pablo de Olavide}, \orgaddress{\city{Sevilla}, \postcode{3800708}, \country{Spain}}}

\affil[5]{\orgdiv{Centro de Ciencias Exactas},  \orgname{Universidad del B\'io-B\'io}, \orgaddress{\city{Chill\'an}, \postcode{3800708},  \country{Chile}}}
\abstract{
In this study, we present a perturbative analysis of the three-gluon vertex for a kinematical symmetric configuration in dimensions $n=4-2\epsilon$ and different covariant gauges. Our study can describe the form factors of the three gluon vertex in a wide range of momentum.
We employ a momentum subtraction (MOM) scheme to define the renormalized vertex. 
We give an in-depth review of three commonly used vector representations  for the vertex, and explicitly show the expressions to change from one representation to the other.
Although our estimates are valid only in the perturbative regime,
we extend our numerical predictions to the infrared domain and show that in $n=4$ some nonperturbative properties are qualitatively present already at perturbation theory.
In particular, we find a critical gauge above which the leading form factor displays the so-called zero crossing.
We contrast our findings to those of other models and observe a fairly good agreement.
}
\maketitle
\section{Introduction}
After half a century of its formulation~\cite{PhysRevLett.30.1343,PhysRevLett.30.1346}, Quantum Chromodynamics (QCD) is nowadays firmly established to be the theory that describes the strong interaction among
quarks and gluons. Its understanding is closely related to the knowledge of vertices involving these two particles.
 Unlike other theories, the elementary particles that mediate the strong interactions, namely, the gluons (discovered at DESY~ \cite{Barber:1979yr,TASSO:1979zyf,Ellis:1976uc} in 1979),  posses color charge and, therefore, can self-interact.
Such an interaction turns out to be relevant even in perturbation theory where e.g. the one-loop expansion for the quark-gluon vertex is defined in terms of the tree level expression for the vertex involving three gluons~\cite{PhysRevD.63.014022}. The quark-gluon vertex has been nonperturbatively studied by lattice QCD~\cite{Skullerud:2001aw,Kizilersu:2006et,Oliveira:2016muq,Kizilersu:2021jen,Oliveira:2018fkj} and Schwinger-Dyson Equations (SDE)~\cite{Davydychev:1996pb,Aguilar:2023mam, Sultan:2018tet,Bermudez:2017bpx,Oliveira:2018ukh,Williams:2014iea}. A comprehensive computation of the twelve form factors linked to this vertex was released at the perturbative level in~\cite{Bermudez:2017bpx}, in all cases for a limited set of kinematical configurations, precisely the symmetric configuration. Slavnov-Taylor identities  (STIs) and Transverse Slavnov-Taylor identities  that constrain the transverse quark-gluon vertex were studied in~\cite{Albino:2018ncl,Albino:2022gzs}.
On the other hand, the three-gluon vertex has 
been the focus of several investigations, as it reflects the non-Abelian nature of QCD 
and is related to nonperturbative phenomena such as confinement, chiral symmetry breaking, and bound-state
formation~\cite{Binosi:2017rwj,Blum:2015lsa,Eichmann:2014xya,Binger:2006sj}.\\
Since the three-gluon vertex is composed of a total of 14 form factors (10 longitudinal and 4 transverse components), which are complicated functions of three independent momenta $p$, $k$, and $q$, it is challenging to study from a technical point of view~\cite{Ball:1980ax}. 
One of the ways to reduce these technical difficulties is the so-called planar degeneracy, which consists of reducing the kinematic dependence of the vertex to a single variable $s^2=1/2(p^2+k^2+q^2),$ leading to an exceptionally straightforward parametrization~\cite{Aguilar:2023qqd,Pinto-Gomez:2022brg} which defines a plane in the coordinate system $(p,k,q)$.
Nevertheless, the fact that lattice simulations can only compute transverse projections of the interaction vertices presents a difficulty in the search to validate the existence of the Schwinger mechanism in QCD~\cite{Skullerud:2003qu,Cucchieri:2006tf,Athenodorou:2016oyh,Boucaud:2017obn,Aguilar:2023qqd,Pinto-Gomez:2022brg}. 
\\
The ‘zero crossing’ property of a form factor
is the transition from positive values to a negative divergence at the origin~\cite{Athenodorou:2016oyh}.
In two- and three-dimensions~\cite{Cucchieri:2008qm}, the leading tree-level component of the three-gluon vertex has a zero crossing at some infrared (IR) momentum scale in a lattice calculation. In 4-dimensions, a similar characteristic is studied in~\cite{Duarte:2016ieu,Athenodorou:2016oyh}.
Its presence, and particularly its location, may have far-reaching consequences for a wide range of hadronic observables: excited states, gluonic components of exotic mesons, hybrids, and glueballs are some of the applications beyond the rainbow-ladder truncation~\cite{Fischer:2009jm}. 
Lattice studies predict that in the symmetric configuration, form factors of the three-gluon vertex exhibits a zero crossing in the IR region around $0.1-0.2$ GeV, below which the data appears to show some form of diverging behavior~\cite{Athenodorou:2016oyh}.
The theoretical basis for this specific feature has been also examined within the context of SDEs~\cite{Eichmann:2014xya,Blum:2015lsa}.\\
The structures of the three-gluon vertex have been studied within the Curci-Ferrari Model (CFM) in~\cite{Pelaez:2006nj}, a nonperturbative implementation of the Ball-Chiu construction (NP-BCC) in~\cite{Aguilar:2019jsj}, and lattice simulations in~\cite{Aguilar:2021lke,Athenodorou:2016oyh,Pinto-Gomez:2022brg,Papavassiliou:2022umz,Parrinello:1994wd}. The three-gluon vertex at one loop in the perturbative regime, in arbitrary dimensions and gauge, was calculated in the article in~\cite{Davydychev:1996pb}.
Studying this vertex in various gauges at one-loop is one of the primary goals of this work; we pay particular attention to the Landau, Feynman, Arbuzov, and Yennie gauges. While the Feynman and Landau gauges are the preferred choice for the sake of simplifying computations, both the Arbuzov and Yennie have demonstrated valuable properties in the ultraviolet and infrared regimes. Studies concur that the Yennie gauge controls ultraviolet behavior while the Arbuzov gauge correctly describes the infrared region of the gluon propagator~\cite{Arbuzov:1980rm,Arbuzov:1982sz,Arbuzov:1986xu,Arbuzov:1987be,Davydychev:1996pb, Yennie:1961ad}. 
The Yennie gauge is commonly utilized in computations within standard Quantum Electrodynamics (QED), but more recently, it has become necessary to utilize it within the context of pseudo-QED in  \cite{Mizher:2024zag}.\\
Additionally, an analysis of the critical point configuration in QCD \cite{Gracey:2023unc,Gracey:2023sup}, by considering the gauge parameter alongside the gauge coupling, suggests that the Arbuzov gauge appears as a stable fixed point, particularly in the infrared region.\\
The remaining of the article is organized as follows. In Sect.~\ref{sec:Ingredients} we describe ingredients and general considerations to compute the form factors of the three-gluon vertex.
In Sect.~\ref{basis}, we present an analysis of the different basis employed herein. In Sect.~\ref{nemerical}, we illustrate our results compared with other theoretical predictions and with lattice simulations in all momenta domain. At the end of the section, we summarize our results. Finally, Sect.~\ref{Conclusions} is devoted for the presentation of our conclusions.
\section{Ingredients and General Considerations}
\label{sec:Ingredients}
In this Section, we set up the nomenclature and describe the roles played by the Green functions in the three-gluon vertex. Additionally, we define the notation that is employed in the integrations that arise during the computation of its one-loop correction.
\begin{itemize}
\item We denote the quark propagator with total momentum $p$ as $\mathcal{S}(p)$.
Two scalar functions 
$\alpha(p^2)$ and $\beta(p^2)$ define the 
inverse quark propagator via 
\begin{equation}
\mbox{i} \mathcal{S}^{-1}(p) \equiv \alpha(p^2)\pslash + \beta(p^2) I \; , 
\end{equation}
where $\pslash\equiv p^{\mu}\gamma_{\mu}$, whereas $I$ is the unit matrix.
\item The  expression for the gluon propagator with momentum $q$ is
\bea
\label{gl_prop}
D_{\mu\nu}^{ab}(q)=-i\delta^{a b} \; \Delta(q)\mathcal{P}_{\mu\nu}(q)\;,
\eea
where $\Delta(p)$ is the gluon dressing function and 
\bea
P_{\mu\nu}(q)=\left( g_{\mu \nu} - \xi \; \frac{q_{\mu} q_{\nu}}{q^2} 
\right)\;.
\eea
The lowest-order propagator is
\begin{equation}
\label{gl_prop2}
D_{\mu\nu}^{ab(0)}(q)=-i\delta^{a b} \; \frac{1}{q^2} 
\left( g_{\mu \nu} - \xi \; \frac{q_{\mu} q_{\nu}}{q^2} 
\right) ,
\end{equation}
where $\xi$ is the covariant gauge parameter defined such that $\xi=0$ is the Feynman gauge and $\xi=1$ is Landau gauge. Other choices of gauges shall be discussed below. $a,b$ are color indices and $p$ is the gluon momentum.\\
In our analysis, we investigate the behavior of the gluon propagator in some particular gauges using the equation introduced in \cite{Boos1988} 
\begin{equation}
\label{gl_prop-g}
D_{\mu\nu}^{ab(0)}(q)=-i\delta^{a b} \; \frac{1}{q^2} 
P_{\mu\nu}(q)+ f(q^2)\frac{q_{\mu} q_{\nu}}{q^2} \;,
\end{equation}
where $f(q^2)$ corresponds to gauge-fixing term. 
\item 
The ghost propagator with mpmentum $p$ is
\begin{equation}
\label{gh_se}
\widetilde{D}^{a b}(p) = i\delta^{a b}\;\frac{G(p^2)}{p^2}\;.
\end{equation}
The lowest-order for the dressed function is $G^{(0)}=1$. In the next Section, we discuss the implications of the function $G$ at the one-loop level in some gauges.
\item  Following~\cite{Davydychev:1996pb}, we define two totally symmetric
combinations of the invariants formed from the external momenta,
\bea
\label{qqq}\nn 
{\cal{Q}} &\equiv & (p k) + (p q) + (k q)= - \frac{1}{2} (p^2 + k^2 + q^2) ,
\\ \nn
{\cal{K}} & \equiv & p^2 k^2 - (p k)^2= p^2 q^2 - (p q)^2= k^2 q^2 - (k q)^2 \,.
\eea
The structure $-4{\cal{K}}$ is
the K\"allen function of $p^2, k^2$ and $q^2$~\cite{Davydychev:1996pb}.
\item Next, we introduce the notation
for the integrals appearing in the one-loop calculations.
The integral involving three points is given by the expression
\begin{equation}
\label{defJ}
J (\mu  ,\nu  ,\alpha) \equiv \int
 \frac{\mbox{d}^n q}{ ((k -q )^2)^{\mu}  ((p +q )^2)^{\nu}
      (q^2)^{\alpha} } ,
\end{equation}
where $n$ is the space-time dimension. In the following Section, we will use $n=4-2\varepsilon$ for 4 dimensions.
When $\mu=\nu=\alpha=1$, the integral simplifies to
\begin{equation}
\label{J(1,1,1)}
J(1,1,1) = {\mbox{i}} \pi^{n/2} \;
\eta \;
\varphi(p^2,k^2,q^2) ,
\end{equation}
where $\varphi(p^2,k^2,q^2) \equiv \varphi$ is a totally
symmetric function.
The integrals with one of the arguments equal to zero, namely 
 $J(0,1,1)$, $J(1,0,1)$ and $J(1,1,0)$, can be written as
\bea
\label{kappa}
\kappa(p^2)&= - \frac{2}{(n-3) (n-4)} \; (-p^2)^{(n-4)/2}= \frac{1}{\varepsilon (1-2\varepsilon)} \; (-p^2)^{-\varepsilon}\;,
\eea
Therefore, we have that
\begin{eqnarray}
J(1,1,0) &= {\mbox{i}} \pi^{n/2} \;
\eta \;
\kappa(q^2) ,\\
J(1,0,1) &= {\mbox{i}} \pi^{n/2} \;
\eta \;
\kappa(k^2),\\
J(0,1,1) &= {\mbox{i}} \pi^{n/2} \;
\eta \;
\kappa(p^2) ,
\end{eqnarray}
$\eta$ denotes a factor constructed of gamma functions, $\Gamma(z)$,
\begin{eqnarray}
\label{eta}
\eta &\equiv& 
\frac{\Gamma^2(\frac{n}{2}-1)}{\Gamma(n-3)} \; 
     \Gamma(3-{\frac{n}{2}})=
\frac{\Gamma^2(1-\varepsilon)}{\Gamma(1-2\varepsilon)} \; 
\Gamma(1+\varepsilon) .
\end{eqnarray}
\end{itemize}
The aforementioned expressions allow us to analyze the three-gluon vertex in the next Section.
\section{Three-Gluon Vertex}
\label{basis}
\begin{figure}[htb]
\vspace{-5cm}
       \centerline{
       \includegraphics[scale=0.5,angle=0]{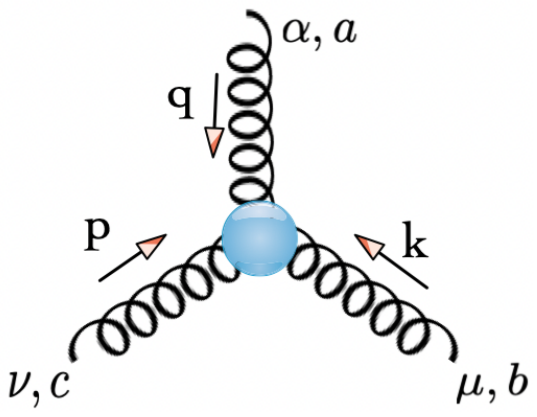}
       }
          \vspace{-5cm}
       \caption{\label{vec-3g} Kinematic configuration for the three-gluon vertex. $a$, $b$, and $c$ are color indices. $\alpha$, $\nu$ and $\mu$, are Lorentz indices. The sum of the incoming momenta is $p~+~k~+~q~=~0.$  }
\end{figure}

\begin{figure}[htb]
\vspace{0cm}
       \centerline{
       \includegraphics[scale=0.5,angle=0]{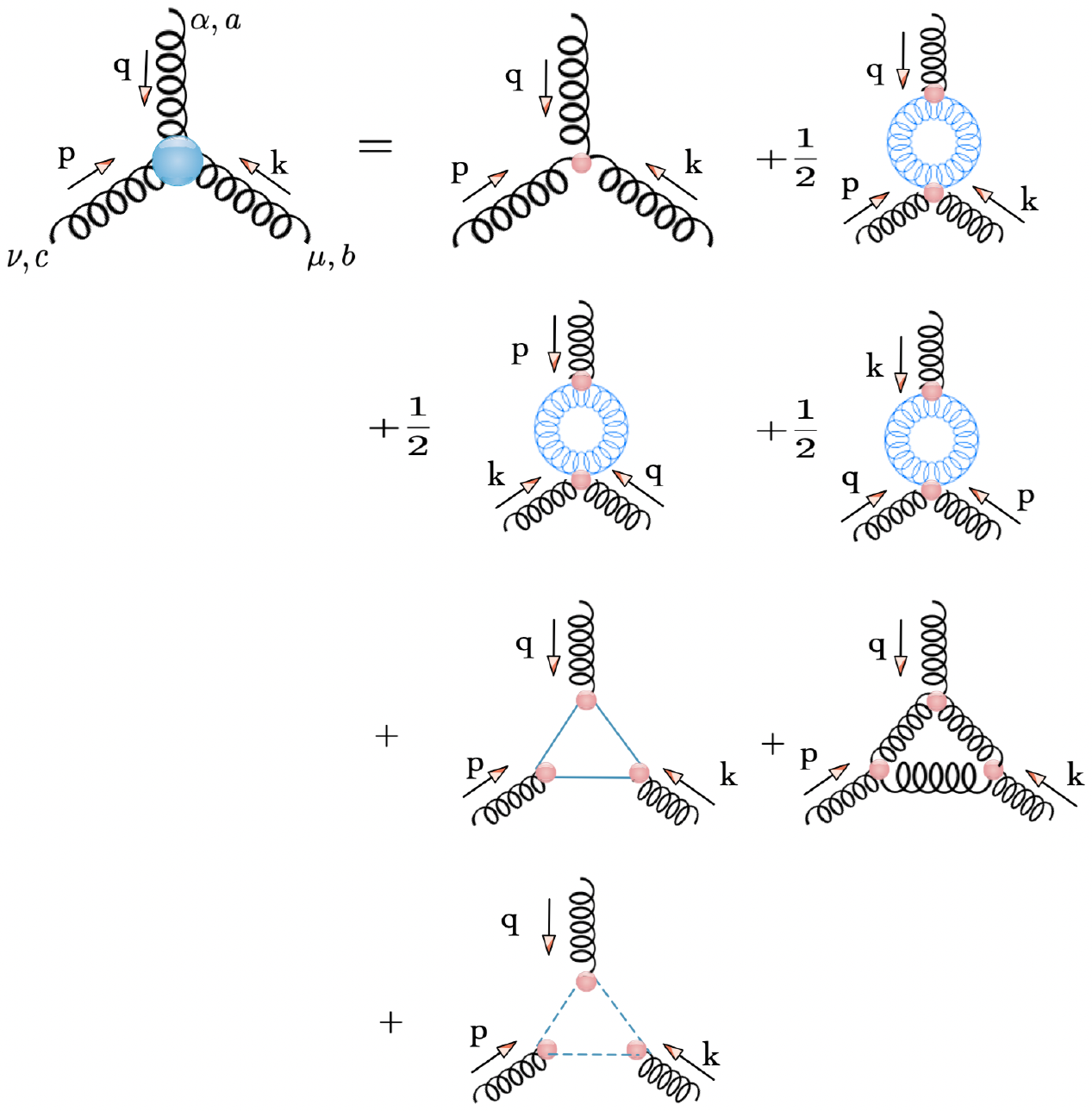}
       }
          \vspace{0cm}
       \caption{\label{one loop-3g}  One-loop corrections to the three-gluon vertex.  Wavy lines represent gluons, continuous lines stand for quarks and dashed lines represent ghosts propagators.}
\end{figure}

One of the fundamental pieces that enlightens us about the non-abelian nature of QCD is the three-gluon vertex. Its most general form can be expressed as \cite{Davydychev:1996pb}
\begin{equation}
\label{ggg}
\Gamma_{\nu \mu \alpha}^{a b c}(p, k, q)
\equiv  - \mbox{i} \; g \;
f^{a b c} \; \Gamma_{
\nu \mu \alpha}(p, k, q) .
\end{equation}
Since gluons are bosons and their color structures 
$f^{abc}$ are antisymmetric, the representation of the moments is as shown in \fig{vec-3g}. 
The lowest-order three-gluon vertex is
\begin{eqnarray}
\label{2eq:tree-level} \nn
\Gamma_{\nu \mu \alpha}^{a b c,(0)}(p, k, q)=
-\mbox{i} \; g \; f^{a b c} 
\bigg[ g_{\nu \mu} (p - k)_{\alpha}+ g_{\mu \alpha} (k - q)_{
\textcolor{blue}{\nu}}
      + g_{\alpha \nu} (q - p)_{\mu}
\bigg]\;. 
\end{eqnarray}
\fig{one loop-3g} depicts radiative corrections to the three-gluon vertex at the one-loop level.
By conservation of momentum, we have that the three incoming momenta 
$p+k+q=0$, and therefore the vertex depends on only two momenta. We have two independent momenta, with three Lorentz indices that, when combined with the metric, give us fourteen structures.
We use the standard notation $C_A$ for the Casimir constant.~\footnote{ $C_F$ and $C_A$ denote eigenvalues of the quadratic Casimir operator in
the fundamental and adjoint representations, respectively.
The diagram in \fig{vec-3g} is non-Abelian; in this case, $C_F$ does not appear.}
\begin{equation}
\label{C_A}
f^{acd}f^{bcd} = C_A \, \delta^{ab} \hspace{5mm} 
(C_A = N \; \mbox{for the SU($N$) group}),
\end{equation}
Through the pertinent SDEs and the gauge symmetry relations of QCD, specifically the Slavnov-Taylor identities (STIs)~\cite{Slavnov:1972fg,Taylor:1971ff}, the vertex $\Gamma_{\alpha\mu\nu}$ is closely related to the corresponding behavior of the gluon and ghost propagators and satisfies, when contracted
with $k^{\mu}$, $q^{\alpha}$, or $p^{\nu}$, the following relations
\bea
\nn q^\alpha\Gamma_{\alpha\mu\nu}(p,k,q) &=& G(q)[\Delta^{-1}(p) P^{\alpha}_\nu(p)\mathcal{H}_{\alpha\mu}(p,q,k)- \Delta^{-1}(k)P^{\alpha}_\mu(k)\mathcal{H}_{\alpha\nu}(k,q,p)] \,, \nonumber\\
\nn k^\mu\Gamma_{\alpha\mu\nu}(p,k,q) &=& G(k)[\Delta^{-1}(q) P^{\mu}_\alpha(q)\mathcal{H}_{\mu\nu}(p,k,q)
- \Delta^{-1}(p)P^{\mu}_\nu(p)\mathcal{H}_{\mu\alpha}(p,k,q)] \,, \nonumber\\
\nn p^\nu\Gamma_{\alpha\mu\nu}(p,k,q) &=& G(p)[\Delta^{-1}(k) P^{\nu}_\mu(k)\mathcal{H}_{\nu\alpha}(k,p,q) - \Delta^{-1}(q)P^{\nu}_\alpha(q)\mathcal{H}_{\nu\mu}(q,p,k)] \,.
\label{eq:sti_delta}
\eea
%
The one-loop ghost dressing function reads
\begin{equation}
\label{G(1)}
G^{(1)}(p^2) = \frac{g^2 \; \eta}{(4\pi)^{n/2}} \;
 \frac{C_A}{4}
\left[ 2 + (n-3)\xi \right] \; \kappa_0(p^2)  ,
\end{equation}
where $g$ is the “re-scaled” coupling constant~\footnote{The value $g$ is a constant corresponding to the running coupling of the theory defined at the renormalization scale.} and $\eta$ is defined in \eqn{eta}.
It is important to note that in the Yennie gauge (corresponding to the covariant gauge parameter $\xi=-2$) $G^{(1)}(p^2)$ is finite as $n\to 4$.  Then, in this gauge, there is no ghost contribution~\cite{Boos1988}.\\ 
$\mathcal{H}_{\mu\nu}(p,k,q)$ appearing in the above STIs stands for 
the ghost-gluon scattering kernel, whose general Lorentz decomposition is given in~\cite{Ball:1980ax,Davydychev:1996pb}~\footnote{The notation $h_i \equiv h_i(p,k,q)$, namely, the momenta dependence of the form factors has been suppressed for compactness.} 
\be\label{eq:H}
\mathcal{H}_{\mu\nu}(p,k,q) = g_{\mu\nu}h_1 + q_\mu q_\nu h_2 + k_\mu k_\nu h_3 + q_\mu k_\nu h_4 + k_\mu q_\nu h_5\,,
\ee
Observe that $\mathcal{H}^{(0)}_{\nu\mu}(p,k,q) = g_{\mu\nu}$ at the tree level, resulting in $h^{(0)}_1 = 1$ and $h^{(0)}_i = 0$ for $i=2,\ldots,5$. For the construction at hand, the nonperturbative structure of the form factors $h_i$ is crucial, and it has been thoroughly examined in~\cite{Aguilar:2018csq}. \\
In the ensuing subsections, we examine the relationships among three widely used representations used to span $\Gamma_{\nu\mu\alpha}$. 
The first is the well-known Ball-Chiu (BC) basis~\cite{Ball:1980ax}, the second one was introduced in~\cite{Davydychev:1996pb} by Davydychev, Osland, and Tarasov (DOT), whereas the third option discussed in this article was introduced in~\cite{Aguilar:2019jsj} by Aguilar, Ferreira, Figueiredo and Papavassiliou (AFFP), and is suitable for lattice calculations.
\subsection{BC Basis}
\label{BC base}
A general form to express the three-gluon vertex was proposed in~\cite{Ball:1980ax} and comprises of longitudinal and transverse components
\bea
\label{BC-ggg} \Gamma_{\nu \mu \alpha}(p, k, q)= \Gamma_{\nu \mu \alpha}^L(p, k, q)+ \Gamma_{\nu \mu \alpha}^T(p, k, q)\;.
\eea
The longitudinal part which contains 10 tensors can be written as~\footnote{We use the notation $(pk)=p\cdot k= |p||k|\cos \theta$.}
\bea
\label{BC-ggg-L}
\nn  \Gamma_{\nu \mu \alpha}^L(p, k, q)&=&
 A(p^2, k^2; q^2)\; g_{\nu \mu} (p-k)_{\alpha}
+ B(p^2, k^2; q^2)\; g_{\nu \mu} (p+k)_{\alpha}
\\ \nn &-& C(p^2, k^2; q^2)\;
\bigg( (p k) g_{\nu \mu} - {p}_{\mu} {k}_{\nu} \bigg)
\; (p-k)_{\alpha}
+ \frac{1}{3}\; S(p^2, k^2, q^2)
\; \bigg( {p}_{\alpha} {k}_{\nu} {q}_{\mu}
        + {p}_{\mu} {k}_{\alpha} {q}_{\nu} \bigg)
\nonumber \\
\nn &+& \bigg\{ \; \mbox{cyclic permutations of} \; (p,\nu),
(k,\mu), (q,\alpha)\; \bigg\}\;.\\
\eea
Since the vertex must satisfy the Ward identity, the transverse part of the
vertex is given in terms of 4 tensors as follows:
\bea
\label{BC-ggg-T}
\nn  && \hspace{-0.5cm} \Gamma_{\nu \mu \alpha}^T(p, k, q)=
 F(p^2, k^2; q^2)\;
\bigg( (p k) g_{
\nu \mu} - {p}_{\mu} {k}_{
\nu} \bigg)
\; \bigg( {p}_{\alpha} (k q) - {k}_{\alpha} (p q) \bigg)
\nonumber \\
\nn &+& H(p^2, k^2, q^2)
\bigg[ -g_{
\nu \mu}
\bigg( {p}_{\alpha} (k q) \!-\! {k}_{\alpha} (p q) \bigg)
+\frac{1}{3}
\bigg( {p}_{\alpha} {k}_{
\nu} {q}_{\mu}
\!-\! {p}_{\mu} {k}_{\alpha} {q}_{
\nu} \bigg) \bigg]
\nn \\
\nn &+& \bigg\{ \; \mbox{cyclic permutations of} \; (p,
\nu),
(k,\mu), (q,\alpha)\; \bigg\} \;,\\
\eea
where the $H$ function is fully symmetric, the $A$, $C$, and $F$ functions are symmetric under the interchange of the first two momenta, the $S$ function is antisymmetric with under the interchange of any pair of arguments, and, finally, the $B$ function is antisymmetric with regard to the first two arguments.
$F$ and $H$ are referred as transverse form factors as the correponding tensor structure vanishes when contracted with
any of ${p}^{\nu}$, $k^{\mu}$ or $q^{\alpha}$.
In the limit $n\to 4$ ($\varepsilon\to 0$), the only 
function which may have an ultraviolet singularity is  $A$, 
since it is the only function which does not vanish at the 
tree level.
\subsubsection{Renormalization}

The longitudinal form factors are defined through
\eqn{BC-ggg-L}.
 $A(p^{2},k^{2},q^{2})$ is the only one of these which
 is UV-divergent at one-loop. 
 We regulate such function with the dimension $n=4-2 \epsilon$ as $\epsilon\to 0$ and employ the momentum subtraction (MOM) renormalization scheme to
 define the renormalized vertex (identified by the subscript $R$ below), such that at a large enough momentum
 scale $p^2=-\mu^2$, tree-level perturbation theory is valid and hence, for the
 symmetric case,
 \bea
   \Gamma^{\mu}_R(p^{2},p^{2},p^{2})\Bigg|_{p^{2}= -\mu^{2}} \equiv \Gamma^{\mu}_R(p^{2},-\mu^2)\Bigg|_{p^{2}= -\mu^{2}} \,.
 \eea
 This renormalization condition translates as:
 \begin{eqnarray}
  A_{1R}(p^{2},p^{2},p^{2})\Bigg|_{p^{2}=-\mu^{2}} \equiv A(p^{2},-\mu^2)\Bigg|_{p^{2}=-\mu^{2}} =
  0 \,,
 \end{eqnarray}
 and determines the vertex renormalization constant $Z_{1F}^{-1}(\mu^{2},\varepsilon)$ as
 follows:
 \begin{eqnarray}
 \Gamma^{\mu}_R(p^{2},-\mu^2)=Z_{1F}^{-1}(\mu^{2},\varepsilon)\Gamma_B^{\mu}(p^2,\varepsilon)
 \,,
 \label{realization}
 \end{eqnarray}
 where the subscript $B$ specifies the bare quantities.
 In one-loop perturbation theory,
\begin{eqnarray}
Z_{1F}(\mu^{2},\varepsilon)=1+A(-\mu^{2},\varepsilon)
\,, \label{renormalization cte}
 \end{eqnarray}
where the bare quantities depend upon the momentum scale $p^2$ and
on the regulator $\epsilon$, having pole divergence
$1/\epsilon$ for $\epsilon \rightarrow 0$. We convert the bare
coupling into the renormalized one through the prescription: $
{g^2}/{4 \pi} = \alpha(\mu) \, {\cal Z_{\alpha} }(\mu^2,
  \epsilon).$
Note that as $
  {\cal Z_{\alpha} }(\mu^2, \epsilon) = 1 + {\cal O}(\alpha)$,
 we can write $
  {g^2}/{4 \pi} = \alpha(\mu)$ at the one-loop order.
\subsection{DOT Representation}
In this Section we analyze the decomposition of the three-gluon vertex proposed in~\cite{Davydychev:1996pb}.
The vertex $\Gamma_{\nu\mu\alpha}(q,k,p)$ can be write as:
\begin{equation}
\label{davy-b}
\Gamma_{\nu\mu\alpha}(p,k,q)=\sum_i^{14}Z_{i}\tau^{i}_{\nu\mu\alpha}\;.
\end{equation}
If we express $q$ in terms of the two other momenta,
$q=-p-k$, we get the following decomposition  in terms of
$Z_{i}$, which are are scalar functions depending on $p, k$ and
$q$~\cite{Davydychev:1996pb}
\bea
 \nn \Gamma_{\nu \mu \alpha}(p, k, q) &=&  g_{\nu \mu} {p}_{\alpha} Z_{1}
 + g_{\nu \alpha} {p}_{\mu} Z_{2}
 + g_{\mu \alpha} {p}_{\nu} Z_{3}
+ g_{\nu \mu} {k}_{\alpha} Z_{4}
 + g_{\nu \alpha} {k}_{\mu} Z_{5}
 + g_{\mu \alpha} {k}_{\nu} Z_{6}\\ \nn
&+& {p}_{\nu} {p}_{\mu} {p}_{\alpha} Z_{7}
+ {k}_{\nu} {k}_{\mu} {k}_{\alpha} Z_{8} 
+ {p}_{\nu} {p}_{\mu} {k}_{\alpha} Z_{9}
+ {p}_{\nu} {k}_{\mu} {p}^{\alpha} Z_{10}
+ {k}_{\nu} {p}_{\mu} {p}_{\alpha} Z_{11}\\ \nn
& +& {p}_{\nu} {k}_{\mu} {k}_{\alpha} Z_{12}
+ {k}_{\nu} {p}_{\mu} {k}_{\alpha} Z_{13}
+ {k}_{\nu} {k}_{\mu} {p}_{\alpha} Z_{14} ,
\eea
where we identify the tensors $\tau^i_{\nu\mu\alpha}$, $i=1,2,\dots, 14$ as
\bea \nn
\begin{array}{@{\extracolsep{0.7cm}}llll}
\tau^1_{\nu\mu\alpha} = g_{\nu \mu} {p}_{\alpha}\;, &
\tau^2_{\nu\mu\alpha} = g_{\nu \alpha} {p}_{\mu}\;, &
\tau^3_{\nu\mu\alpha} =  g_{\mu \alpha} {p}_{\nu}\;, &
\tau^4_{\nu\mu\alpha}= g_{\nu \mu} {k}_{\alpha}\;, \\
\rule{0ex}{2.5ex}
\tau^5_{\nu\mu\alpha} = g_{\nu \alpha} {k}_{\mu}\;, &
\tau^6_{\nu\mu\alpha}= g_{\mu \alpha} {k}_{\nu}\;, &
\tau^7_{\nu\mu\alpha}= {p}_{\nu} {p}_{\mu} {p}_{\alpha}\;, &
\tau^8_{\nu\mu\alpha}= {k}_{\nu} {k}_{\mu} {k}_{\alpha}\;, \\
\rule{0ex}{2.5ex}
\tau^9_{\nu\mu\alpha}= {p}_{\nu} {p}_{\mu} {k}_{\alpha}\;, &
\tau^{10}_{\nu\mu\alpha}= {p}_{\nu} {k}_{\mu} {p}_{\alpha}\;, &
\tau^{11}_{\nu\mu\alpha}= {k}_{\nu} {p}_{\mu} {p}_{\alpha}\;, &
\tau^{12}_{\nu\mu\alpha}= {p}_{\nu} {k}_{\mu} {k}_{\alpha}\;, \\
\rule{0ex}{2.5ex}
\tau^{13}_{\nu\mu\alpha}= {k}_{\nu} {p}_{\mu} {k}_{\alpha}\;, &
\tau^{14}_{\nu\mu\alpha}= {k}_{\nu} {k}_{\mu} {p}_{\alpha}\;.
\end{array}
\eea
Although it is apparent that in this representation we cannot not split immediately the vertex into longitudinal and transverse parts, it is nevertheless highly helpful to examine every structure that is a part of the calculation. Of course, we can convert all the structures to a BC form, as described in~\cite{Davydychev:1996pb} and that we address in appendix~\ref{D-BC}.
We also present the formulas to switch from the BC basis to the Davydychev form in appendix~\ref{BC-D}.
The fact that all of the vertex structures have been obtained at one-loop with an arbitrary gauge and dimension gives another benefit of using the DOT representation: it enables us to perform a straightforward numerical study of their behavior. 
\subsection{AFFP Representation}
\label{aguilar basis}
Finally, we use one of the most recent and most used decomposition of the vertex in both Schwinger-Dyson equations and lattice formalisms. This representation  allows us to compare numerically with other methods in different gauges.
The building block of this proposal for the vertex is the BC basis, but with a point of view of a non-perturbative construction and can be written as~\cite{Aguilar:2019jsj}
\begin{equation}
\Gamma_{\nu\mu\alpha}(p,k,q)=\Gamma_{\nu\mu\alpha}^L(p,k,q)+\Gamma_{\nu\mu\alpha}^T(p,k,q)\;,
\end{equation}
where $\Gamma_{\nu\mu\alpha}^L(p,k,q)$ is the longitudinal part and the transverse part $\Gamma_{\nu\mu\alpha}^T(p,k,q)$ satisfies
\begin{equation}
q^\alpha\Gamma_{\nu\mu\alpha}^T(p,k,q)=k^\mu \Gamma_{\nu\mu\alpha}^T(p,k,q)=p^\nu\Gamma_{\nu\mu\alpha}^T(p,k,q)=0\;.
\end{equation}
For the explicit tensorial decomposition of $\Gamma_{\nu\mu\alpha}^L(p,k,q)$ and $\Gamma_{\nu\mu\alpha}^T(p,k,q)$
we  employ the Bose symmetric basis introduced in~\cite{Ball:1980ax}.
Specifically, 
\be
\Gamma_{\nu\mu\alpha}^L(p,k,q) = \sum_{i=1}^{10} X_i(p,k,q) \ell^i_{\nu\mu\alpha} \,,
\label{eq:3g_sti_structure}
\ee
where the tensors $\ell^i_{\nu\mu\alpha}$ are given by
\bea
\begin{array}{@{\extracolsep{0.7cm}}ll}
\nn \ell^1_{\nu\mu\alpha} =  (p-k)_{\nu} g_{\alpha\mu}\,,
& \ell^2_{\nu\mu\alpha} =  - q_{\nu} g_{\alpha\mu}\,,
\\
\nn \ell^3_{\nu\mu\alpha} =  (p-k)_{\nu}[p_{\mu} k_{\alpha} -  (p\cdot k) g_{\alpha\mu}]\,, 
& \ell^4_{\nu\mu\alpha} = (k-q)_{\alpha} g_{\mu\nu}\,,
\\
\nn \ell^5_{\nu\mu\alpha} =  - p_{\alpha} g_{\mu\nu}\,,
& \ell^6_{\nu\mu\alpha} =  (k-q)_{\alpha}[k_{\nu} q_{\mu} -  (k\cdot q) g_{\mu\nu}]\,,
\\
\nn\ell^7_{\nu\mu\alpha} =  (q-p)_{\mu} g_{\alpha\nu}\,,
& \ell^8_{\nu\mu\alpha} = - k_{\mu} g_{\alpha\nu}\,,
\\
\nn\ell^9_{\nu\mu\alpha} = (q-p)_{\mu}[q_{\alpha} p_{\nu} -  (q\cdot p) g_{\alpha\nu}]\,,
& \ell^{10}_{\nu\mu\alpha} = p_{\nu}k_{\alpha}q_{\mu} + p_{\mu}k_{\nu}q_{\alpha}\,, 
\label{li}
\end{array}
\eea
and 
\be
\Gamma_{\nu\mu\alpha}^T(p,k,q) = \sum_{i=1}^{4}Y_i(p,k,q)t^i_{\nu\mu\alpha} \,,
\label{eq:3g_tr_structure}
\ee
with the $t_i^{\alpha\mu\nu}$ given by
\begin{align}
t^1_{\nu\mu\alpha} =& [(p\cdot k)g_{\alpha\mu} - p_{\mu}k_\alpha][(k\cdot q)p_\nu - (p\cdot q)k_\nu]\,,
\nonumber\\
t^2_{\nu\mu\alpha} =& [(k\cdot q)g_{\mu\nu} - k_{\nu}q_\mu][(q\cdot p)k_\alpha - (k\cdot p)q_\alpha]\,,
\nonumber\\
t^3_{\nu\mu\alpha} =& [(q\cdot p)g_{\nu\alpha} - q_{\alpha}p_\nu][(p\cdot k)q_\mu - (q\cdot k)p_\mu]\,,
\nonumber\\
\nn t^4_{\nu\mu\alpha} =& g_{\mu\nu}[ (k\cdot p)q_\alpha - (q\cdot p)k_\alpha ] + g_{\nu\alpha}\bigg[ (q\cdot k)p_\mu -
(p\cdot k)q_\mu \bigg] + g_{\alpha\mu}[ (p\cdot q)k_\nu - (k\cdot q)p_\nu ] 
\\+ &
k_\alpha q_\mu p_\nu - q_\alpha p_\mu k_\nu \,.
\label{ti}
\end{align}
Bose symmetry with respect to the three legs requires that $\GL$ reverses sign under the interchange of the corresponding Lorentz indices and momenta (recall that the color factor $f^{abc}$ has been factored out); this, in turn, imposes the following relations under the exchange of arguments~\cite{Ball:1980ax}
\bea
\begin{array}{@{\extracolsep{1cm}}ll}
X_1(p,k,q) = X_1(k,p,q)\,, &   X_2(p,k,q) = - X_2(k,p,q)
  \\
X_3(p,k,q) = X_3(k,p,q)\,,  &  X_4(p,k,q) = X_4(p,q,k)\,, 
   \\
X_5(p,k,q) = - X_5(p,q,k)\,, &    X_6(p,k,q) = X_6(p,q,k)\,,     
\\
X_7(p,k,q) = X_7(q,k,p)\,, & X_8(p,k,q) = - X_8(q,k,p)\,,
 \nonumber    \\
 X_9(p,k,q)= X_9(q,k,p)\,, & X_{10}(p,k,q) = - X_{10}(k,p,q), \; \\
X_{10}(p,k,q) = - X_{10}(p,q,k), &
\; X_{10}(p,k,q) = - X_{10}(q,k,p). 
\end{array}
\eea
In addition, Bose symmetry furnishes the following relations between different form factors~\cite{Ball:1980ax}
\begin{align}
X_4(p,k,q) &= X_1(k,q,p)\,, \qquad  X_5(p,k,q) = X_2(k,q,p)\,, 
 \nonumber\\
X_6(p,k,q) &= X_3(k,q,p)\,, \qquad X_7(p,k,q) = X_1(q,p,k)\,, 
 \nonumber\\
X_8(p,k,q) &= X_2(q,p,k)\,,  \qquad X_9(p,k,q) = X_3(q,p,k) \,,
\label{eq:Xi_bose}
\end{align}
As in the cases of the BC and DOT representations, we consider the conversion between different decompositions of the vertex.
In appendix~\ref{D-A} we show the expressions to go from the DOT to AFFP representations.
\section{Numerical Results}
\label{nemerical}
Now, we perform a numerical analysis of the results utilizing the three decomposition of the three-gluon vertex discussed above. Considering this, we use the results of ~\cite{Davydychev:1996pb} for arbitrary gauge and dimension.
Our analysis is accomplished in different gauges in the symmetric limit.
For the vertex function to one-loop level at the symmetric point,
\bea
p^2=k^2=q^2 \equiv Q^2\;,
\eea
we get 
$(p k)=(p q)=(k q)=-\frac{1}{2} p^2 
= \frac{1}{2}M^2$.
With these assumptions, the vertex simplifies, and the first observation is that the antisymmetric functions are zero, i.e., $B$ and $S$ functions in the BC basis
\begin{equation}
B(Q^2, Q^2; Q^2) = S(Q^2, Q^2, Q^2) \equiv 0 .
\end{equation}
In the DOT representation, the three-gluon vertex undergoes a significant simplification, reducing from fourteen structures to just three, in the notation used in~\cite{Celmaster:1979km},
as
\begin{eqnarray}
 \nn\Gamma_{\nu\mu\alpha}(p,k,q)&=&
G_0(p^2) \bigg[ g_{\nu\mu}(p-k)_{\alpha}
+ g_{\mu\alpha}(k-q)_{\nu}
+ g_{\alpha\nu}(q-p)_{\mu} \bigg]-
\nonumber \\
&& G_1(p^2) (k-q)_{\nu} (q-p)_{\mu} (p-k)_{\alpha}+ G_2(p^2)
\left( {p}_{\alpha} {k}_{\nu} {q}_{\mu}
-  {p}_{\mu} {k}_{\alpha} {q}_{\nu} \right) ,
\hspace{30mm}
\end{eqnarray}
with the three $G_i$ functions related to
the scalar functions in (\ref{BC-ggg}) through~\footnote{
We have omitted the dependence of the three momenta, since $Q^2$ denotes the single momentum scale available.}
\begin{eqnarray}
\nn \label{gsf2} G_0(Q^2)
&=& A(Q^2) + \frac{1}{2} Q^2 C(Q^2)
+ \frac{1}{4} Q^4 F(Q^2)
+\frac{1}{2}  Q^2 H(Q^2) , 
\hspace*{8mm}
 \\
\nn G_1(Q^2)
&=&C(Q^2) + \frac{1}{2}  Q^2 F(Q^2) ,
\\ 
G_2(Q^2)
&=&C(Q^2) +\frac{1}{2}  Q^2 F(Q^2)+ H(Q^2) .
\end{eqnarray}
We observe that two of these relations can be concisely represented as~\footnote{Although we are not using the  subscript $R$, the function $A$ that we are using is already renormalized.}
\bea
\nn G_2(Q^2)
&=&G_1(Q^2)+H(Q^2) , 
\\
\nn G_0(Q^2)
&=&A(Q^2)+\frac{1}{2} Q^2 G_2(Q^2) .
\eea
In the symmetric limit and one loop for an arbitrary gauge, the expression for  $G_i$ take the form~\cite{Davydychev:1996pb}~\footnote{We denote the one-loop order
contribution with a superscript (1).} 
\begin{eqnarray}
\nn G_0^{(1)}(Q^2) &=& \frac{g^2 \; \eta}{(4\pi^2)} \;
C_A \; \frac{1}{288}
\bigg\{(3\xi(2\xi+5)+2)\left[\mu^2\varphi(\mu^2)-Q^2\varphi(Q^2)\right]
 + 6(9\xi+8)\ln\left(\frac{p^2}{\mu^2}\right)
\bigg\}\label{G0r} ,
\\
\nn G_1^{(1)}(Q^2) &=& \frac{g^2 \eta}{432(4\pi^2)} \;
C_A \; \frac{1}{Q^2}
\bigg\{
 (3 \xi (\xi (4\xi + 3) + 36)-((64 - 3\xi (\xi(2 \xi + 3) - 6)) Q^2 \varphi(Q^2)) - 32)
\bigg\},\\
\label{G1r}
\nn G_2^{(1)}(Q^2) &=& \frac{g^2 \eta}{(4\pi^2)} \;
C_A \; \frac{1}{432 \; Q^2}
\bigg\{ 3\xi (7 \xi (\xi+3)-54) +(3 \xi (\xi (2 \xi+9)-3)+32) Q^2 \varphi(Q^2)-56
\bigg\}.\\
\label{G2r}
\eea
The only divergent function in this basis is $G_0^{(1)}$, illustrated by the \eqn{gsf2}, which depends on $A$ but is not finite at $n=4-2\epsilon$. As a result, $G_0$ is renormalized in \eqn{G2r} using the same formalism as explained in Section~\ref{BC base}.
The integrals in this configuration can be evaluated in terms of Clausen functions as
\begin{eqnarray}\nn
&&J_1(1,1,1)\Bigg|_{n=4,\; Q^2=-\mu^2}
\hspace{-0.8cm}= -\frac{{\mbox{i}}\pi^2}{\mu^2\sqrt{3}}
\bigg\{ 2\Cl{2}{\frac{\pi}{3}} 
+ 2 \Cl{2}{\frac{\pi}{3}+2\theta_{{\rm s}1}}
+ \Cl{2}{\frac{\pi}{3}-2\theta_{{\rm s}1}}
+ \Cl{2}{\pi-2\theta_{{\rm s}1}} \bigg\} \; ,
\end{eqnarray}
\bea \nn
 J_2(1,1,1)\Bigg|_{n=4,\; Q^2=-\mu^2}\hspace{-0.8cm}
&=& -\frac{2{\mbox{i}}\pi^2}{\mu^2\sqrt{3}}
\bigg\{ 2 \Cl{2}{\frac{2\pi}{3}} + \Cl{2}{\frac{\pi}{3}+2\theta_{{\rm s}2}}
+ \Cl{2}{\frac{\pi}{3}-2\theta_{{\rm s}2}} \bigg\} \; ,
\eea
where 
\be
\tan\theta_{{\rm s}1}=\frac{\mu^2+2m^2}{\mu^2\sqrt{3}}, \hspace{10mm}
\tan\theta_{{\rm s}2}=\sqrt{\frac{\mu^2+4m^2}{3\mu^2}} \; .
\ee
In the massless limit ($m\to 0$), $\theta_{{\rm s}1}=\theta_{{\rm s}2}
=\pi/6$, and, recalling that
$\Cl{2}{\frac{2\pi}{3}}=\frac{2}{3}\Cl{2}{\frac{\pi}{3}}$,
we reproduce the well-known result \cite{Celmaster:1979km}
\be
J_0(1,1,1)\Bigg|_{n=4,\; p_i^2=-\mu^2}
= -\frac{4{\mbox{i}}\pi^2}{\mu^2\sqrt{3}} \Cl{2}{\frac{\pi}{3}} \; .
\ee
With all the ingredients discussed above and using $\alpha=0.118$ and the renormalization parameter $\mu=4.3$~GeV, we plot the three form factors of the vertex in the Landau gauge in \fig{gs}. In contrast to $G_1^{(1)}$, which increases, it is clear that $G_0^{(1)}$ and $G_2^{(1)}$ are decreasing in the infrared.
\begin{figure}[htb]
\vspace{0 cm}
       \centerline{
       \includegraphics[scale=0.27,angle=0]{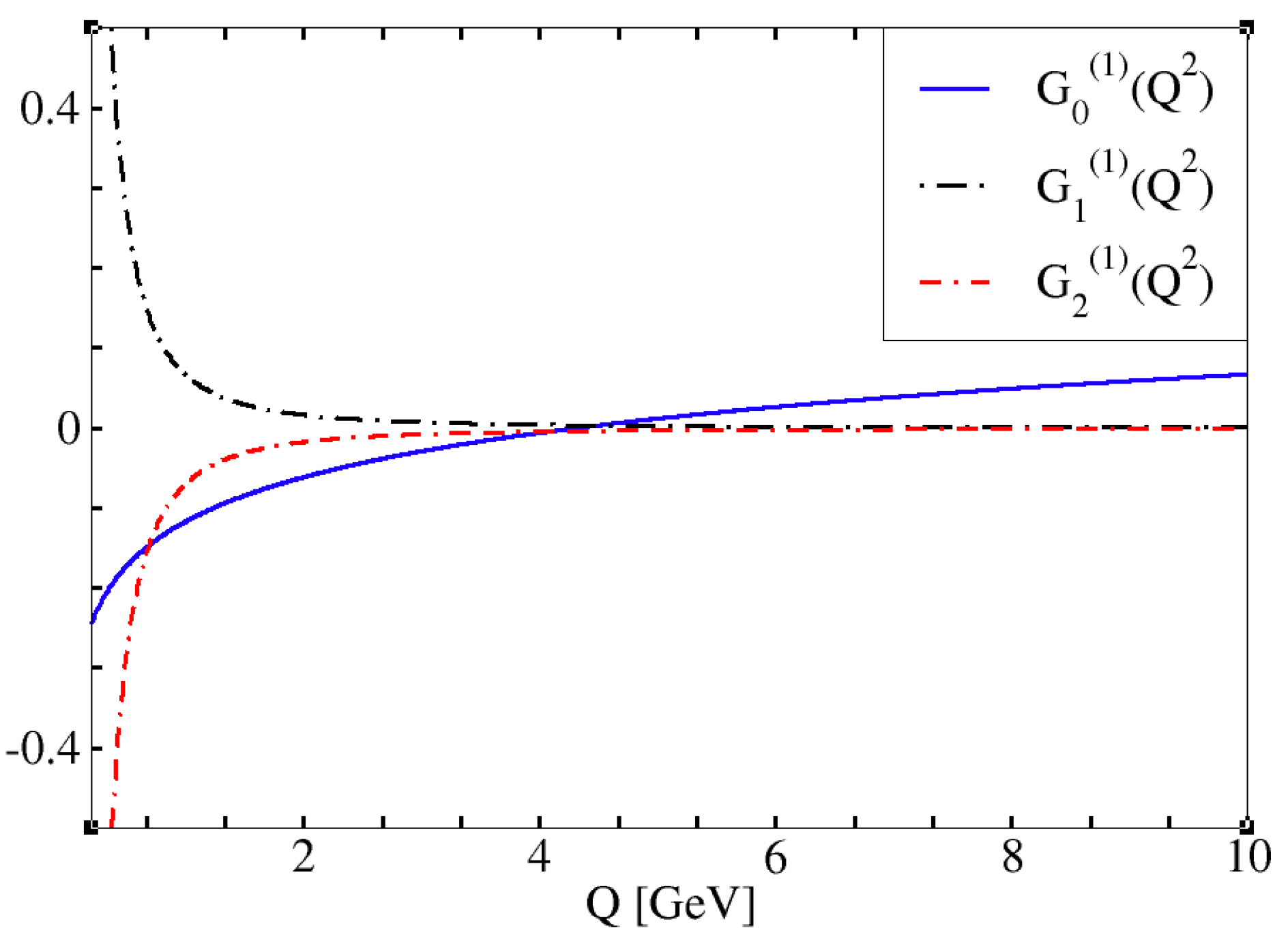}
       }
          \vspace{0cm}
       \caption{\label{gs} Three form factors in \eqn{G2r} of the three-gluon vertex in the symmetric limit in Landau gauge. Although $G_0^{(1)}$ is the only function that has been renormalized, the three curves cross zero close to $Q=4.3$~GeV.  }
\end{figure}
There are no numerical results for these $G_i$ using other methods, but there are for the basis presented in~\cite{Aguilar:2019jsj}. However, as mentioned in the appendix~\ref{D-A}, we can relate these $G$'s with the structures used in the AFFP representation. The relation among the $G$s, $X$s, and $Y$s in Eqs.~(\ref{eq:3g_sti_structure}) and~(\ref{eq:3g_tr_structure}) are~\footnote{ $X_i$ represents
the contribution of the quark loops, $X_i=1+X_i^{(1)}$}
\bea
\nn  X_1(Q)&=&\frac{G_2(Q^2) Q^2}{2}-G_0 (Q^2)\;,
 \;\; \;
X_3(Q)=G_1(Q^2)\;,\\
Y_4(Q)&=&-G_1(Q^2)-G_2(Q^2)\;.
\eea
The only divergent form factor that must be renormalized is $X_1$, because it depends on $G_0$, and hence on $A$.
It is immediately evident that $X_2,X_5,X_8$ and $X_{10}$ in this limit are zero due to their antisymmetric nature, in accordance with the Bose symmetries displayed in subsection (\ref{aguilar basis}).
Additionally, we can observe that the following equalities are satisfied
\bea
\nn X_1(Q)=X_4(Q)=X_7(Q)\;,\\
X_3(Q)=X_6(Q)=X_9(Q)\;.
\eea
The explicit expressions for the components that are non zero in the AFFP representation in arbitrary gauge at one-loop are
\bea
 \nn   X_1^{(1)}(Q)&=&\frac{C_{A} g^2 \eta}{576 \pi^2} \bigg\{-3 (9 \xi+8)\ln\left(\frac{Q^2}{\mu^2}\right)
+(6 \xi^2+9 \xi+17)  \left[-\mu^2\varphi(\mu^2)+Q^2\varphi(Q^2)\right]
\bigg\}\;,
\\
\nn
X_3^{(1)}(Q)&=& -\frac{C_{A} g^2 \eta}{3456 \pi^2 Q^2} \bigg\{12 \xi^3+9 \xi^2+108 \xi+(6 \xi^3+9 \xi^2-18 \xi-64) Q^2 \varphi(Q^2)-32\bigg\}\;,\\
\nn Y_4^{(1)}(Q)&=& \frac{C_A g^2 \eta}{3456 \pi^2 Q^2} \bigg\{3 \xi (\xi (10 \xi]+39)-144)
+(6 \xi^3+45 \xi^2+128) Q^2 \varphi(Q^2)-80\bigg\}\;.
\eea
These formulations are reduced as follows in the Landau gauge $\xi=1$
\bea \nn
X_1^{(1)}(Q)&=&\frac{C_{A} g^2 \eta}{576 \pi^2} \bigg\{ 32 \left[-\mu^2\varphi(\mu^2)+Q^2\varphi(Q^2)\right]
- 51\log\left(\frac{Q^2}{\mu^2}\right)\bigg\}\;,\\
\nn X_3^{(1)}(Q)&=& -\frac{C_A g^2 \eta}{3456 \pi^2 Q^2} (97-67 Q^2 \varphi(Q^2)))\;,
\\ \nn
Y_4^{(1)}(Q)&=&\frac{C_A g^2 \eta} {3456 \pi^2 Q^2}\{179 Q^2 \varphi(Q^2)-365\}\;.
\eea
We plot the expression for $X_1(Q)$  in~\fig{fig-X1} and contrast the outcomes with those found in~\cite{Aguilar:2021lke} using a nonperturbative BC construction of the three-gluon vertex and the one-loop result using a particular version of the general momentum subtraction (MOM)
scheme, known as “Taylor scheme” (TR)~\cite{Aguilar:2019jsj, Grassi:1999tp}. The indistinguishability of the two results at one-loop is instantly apparent. \\
\begin{figure}[htb]
\vspace{-0cm}
       \centerline{
       \includegraphics[scale=0.25,angle=0]{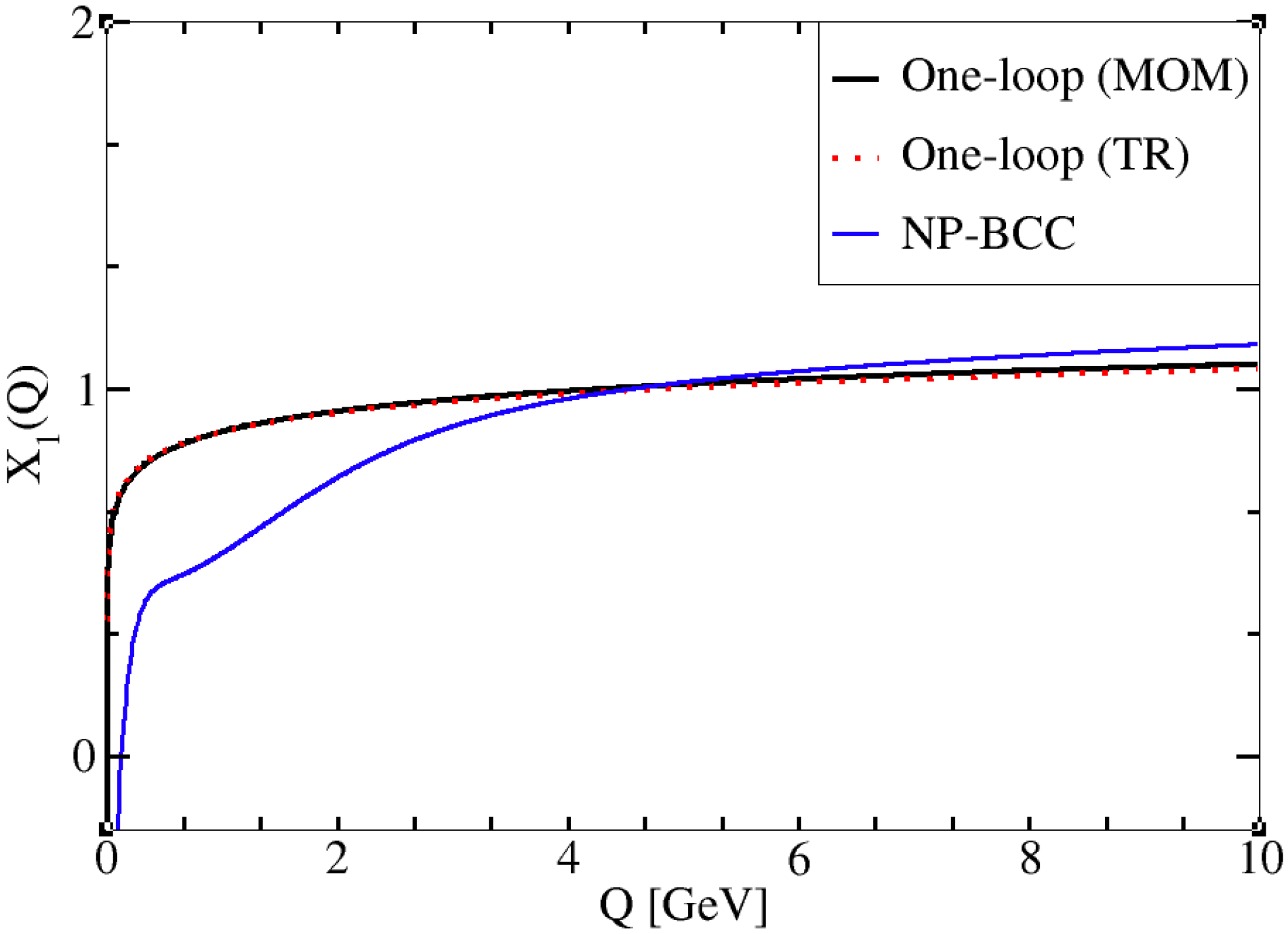}
       }
          \vspace{0cm}
       \caption{\label{fig-X1} $X_1(Q)$ in Landau gauge compared with results in~\cite{Aguilar:2019jsj} using a nonperturbative BC construction (NP-BCC) of the three-gluon vertex. The curves blue and red are plotted with a value $\alpha=0.22$ unlike our result, where $\alpha=0.118$. }
\end{figure}
We can construct a combination of the vertex structures that are non-zero, just as in Refs.~\cite{Aguilar:2019jsj} and~\cite{Aguilar:2021lke}
\bea \label{Lsym-eq}
\nn L_{sym}(Q)&=&X_1(Q)-\frac{Q^2}{2}X_3(Q)+\frac{Q^4}{4}Y_1(Q)-\frac{Q^2}{2}Y_4(Q)\;.
\eea
This expression is plotted in \fig{Lsym} for two different one-loop renormalization schemes.
\begin{figure}[htb]
\vspace{-0.cm}
       \centerline{
       \includegraphics[scale=0.27,angle=0]{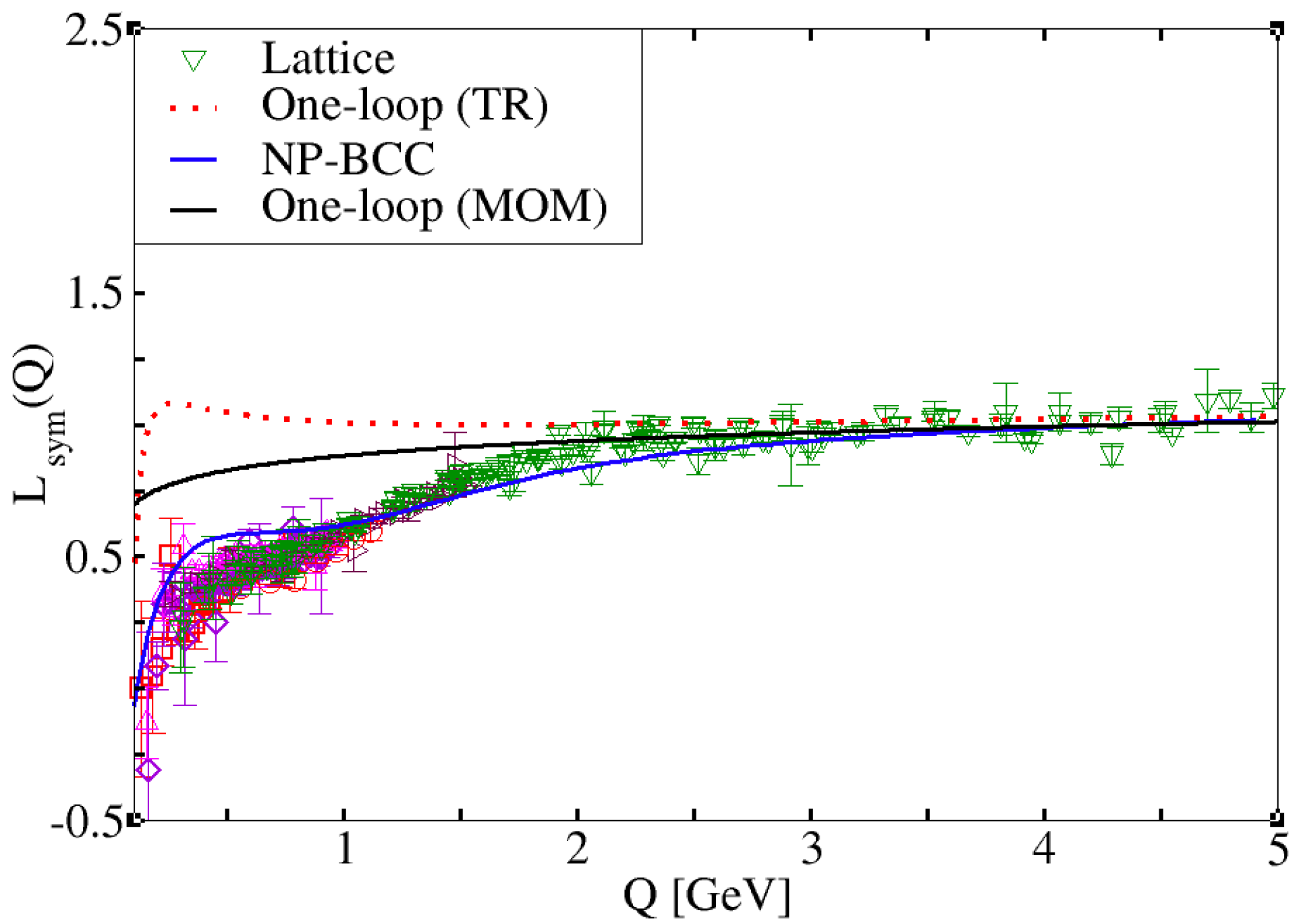}
       }
          \vspace{0cm}
       \caption{\label{Lsym} $L_{sym}(Q)$ in Landau gauge at one-loop with a different renormalization schemes. Lattice results are taken from Refs.\cite{Aguilar:2021lke, Athenodorou:2016oyh}. The blue line corresponds to the result obtained in~\cite{Aguilar:2019jsj}. Results with lattice  (in green inverted triangles) is reported in~\cite{Athenodorou:2016oyh} obtained using the Wilson gauge action at several  bare couplings (ranging from 5.6 to 6.0).}
\end{figure}
\\
It is evident that $X_1$ is the structure that most significantly contributes to $L_{sym}(Q)$  given that
\bea \label{res}
\nn &-&\frac{Q^2}{2}X_3^{(1)}(Q)+\frac{Q^4}{4}Y_1^{(1)}(Q)-\frac{Q^2}{2}Y_4^{(1)}(Q)\\
&=&\frac{C_A g^2 \eta (77-41 Q^2 \varphi(Q))}{1152 \pi^2}\thickapprox 0.0677265 \;,
\eea
the previous expression is given in the Landau gauge.
In \fig{Lsym}, we plot our results in the Landau gauge compared to other theoretical results at one-loop  ~\cite{Aguilar:2019jsj, Grassi:1999tp}, lattice results Refs.~\cite{Aguilar:2021lke, Athenodorou:2016oyh} and other approach using a Nonperturbative BC construction of the three-gluon vertex~\cite{Aguilar:2019jsj}. \\
For physical quantities, gauge invariance can be explicitly monitored by knowing the findings in an arbitrary gauge~\cite{Davydychev:1996pb}.
We want to point out their focus on some particular values of the gauge parameter in the \fig{Lnormas}, where we show the behavior of $L_{sym}(Q)$ for the gauges with $\xi=0,1,2,3,4,-1,-2,-3,-4$. Our results indicate that Feynman and the positive gauges are increasing in the infrared while the negative gauges are falling.
Actually, we identify the critical value $\xi_c=-0.89$ at which $L_{sym}(Q)$ maintains almost constant in the infrared.
The zoom shows that the value
where the curves cross zero increases with increasing the value of the gauge.
\begin{figure}[htb]
\vspace{0cm}
       \centerline{
       \includegraphics[scale=0.43,angle=0]{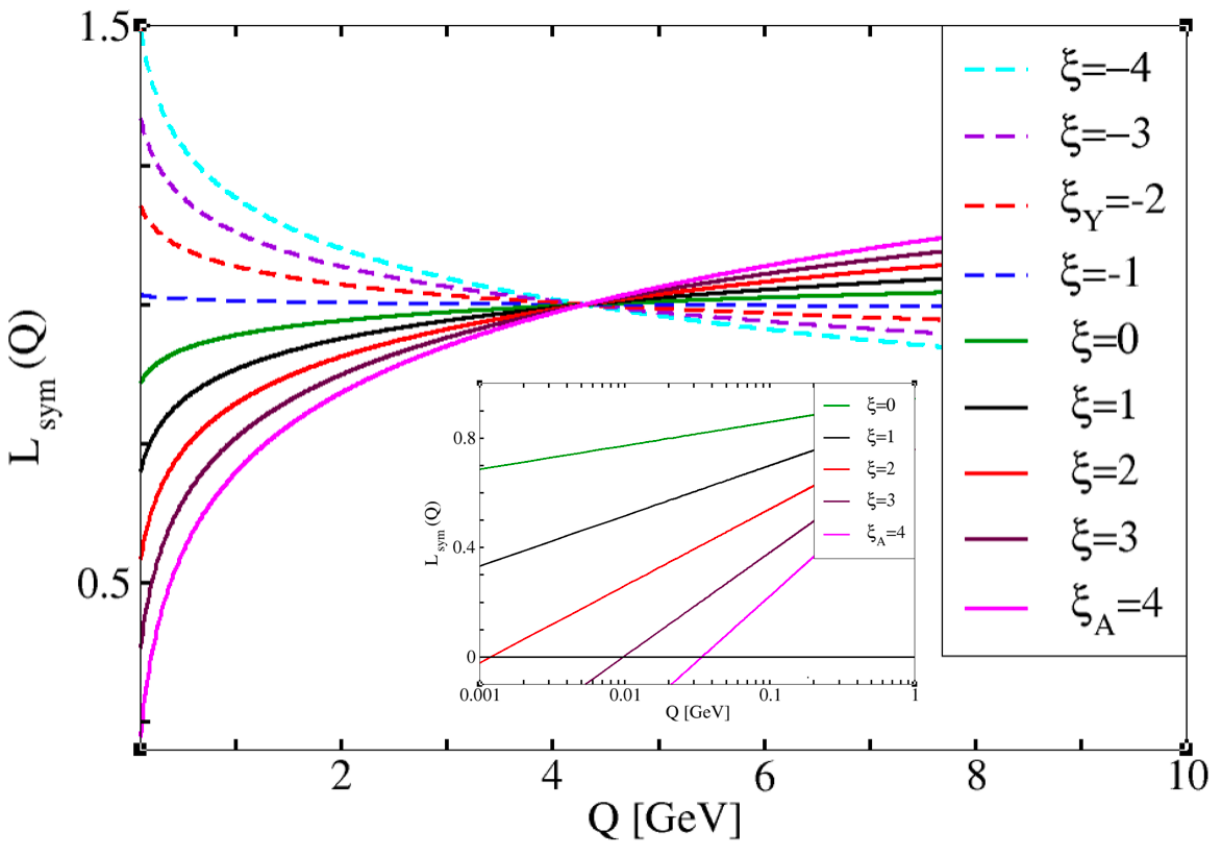}
       }
          \vspace{0cm}
       \caption{$L_{sym}(Q)$ For different gauges. We extracted the result of  Yennie and Arbuzov gauges $\xi_Y=-2$ and $\xi_A=4$, respectively. The renormalization requirement means that at $\mu=4.3$~GeV, all curves are equal to 1. At the bottom of the graph, we display a close-up near $Q=0$. }
       \label{Lnormas}
\end{figure}
Although the whole set of gauge-invariant diagrams that depict the physical processes is gauge-independent, the individual diagrams and the degree of computational complexity are highly dependent on the gauge selection.
Making the right gauge option can make it much easier to calculate radiative corrections in a given situation.
Due to these characteristics, 
we conduct our analysis of the three-gluon vertex using two widely utilized gauges: Yennie and Arbuzov.
\\\\
\textbf{\underline{Yennie Gauge ($\xi_Y=-2)$~:}}
\\\\
Due to its appealing infrared features, the Yennie gauge~\cite{Fried:1958zz} has been found to be helpful for the computation of radiative adjustments to bound-state parameters~\cite{Sapirstein:1983xr}.
Perhaps the Yennie gauge most advantageous technical characteristic is that, as compared to other covariant gauges, it exhibits far better-infrared behavior for individual diagrams. This property is shared by the noncovariant Coulomb gauge as well. Specifically, a great deal of diagrams that are infrared finite in the Yennie gauge but infrared divergent in other covariant gauges~\cite{Eides:2001dw}.
\\\\
\textbf{\underline{Arbuzov Gauge ($\xi_A=4)$~:}}
\\\\
The distinct nature of Arbuzov gauge has already been recognized in the study of the quark propagator in QCD~\cite{Arbuzov:1982sz}. 
The Arbuzov gauge was shown to yield a self-consistent
description of the lowest order gluon and ghost Green’s functions (in other cases, one has to allow
for the non-trivial effect of ghosts on the gauge identities, which are used to reconstruct
the gluon vertices).\\\\
Our findings lend support to those obtained in \cite{Boos1988}, which provide an ansatz for the gluon propagator. This ansatz features a Fourier transform that is transverse in coordinate space and exhibits an arbitrary power-like behavior:

\begin{equation}
\label{gl_prop-g2}
D_{\mu\nu}^{ab}(p)=-i\delta^{a b} \; \frac{(M^2)^{\gamma-1}}{(p^2)^\gamma} 
\left( g_{\mu \nu} -d(\gamma) \frac{p_{\mu} p_{\nu}}{p^2} 
\right)\;,
\end{equation}
where
\bea
d(\gamma)=\frac{2\gamma}{2\gamma+1-d}\;.
\eea
Notice that the
Fourier transform $D_{\mu\nu}^{ab}(x)$ of the propagator\eqref{gl_prop-g2} is transverse:
$D_{\mu\nu}^{ab}(x)x^\nu = 0$. This property ensures the canceling of the principal infrared singularities~\cite{Arbuzov:1982sz}.
In the particular case $\gamma = 1$, \eqn{gl_prop-g2}) corresponds precisely to the
Yennie gauge~\cite{Yennie:1961ad}. When  $\gamma = 2$, it correspond to the Arbuzov gauge~\cite{Arbuzov:1980rm,Arbuzov:1987be,Arbuzov:1986xu}. \\
Consequently, the gluon propagator in \eqn{gl_prop-g2} in both the infrared (IR) and ultraviolet (UV) regions:
\begin{equation}
\label{gl_prop-gIR}
D_{\mu\nu}^{ab(\text{IR})}(p)=-i\delta^{a b} \; \frac{M^2}{(p^2)^2} 
\left( g_{\mu \nu} -d(2) \frac{p_{\mu} p_{\nu}}{p^2} 
\right)\;,
\end{equation}
\begin{equation}
\label{gl_prop-guv}
D_{\mu\nu}^{ab\text{(UV)}}(p)=-i\delta^{a b} \; \frac{1}{p^2} 
\left( g_{\mu \nu} -d(1) \frac{p_{\mu} p_{\nu}}{p^2} 
\right)\;,
\end{equation}
The validity of the \eqn{gl_prop-gIR} is examined in Ref. \cite{Arbuzov:1986xu} using the Schwinger–Dyson equations for the gluon propagator. Moreover, the Arbuzov gauge has been shown to provide an accurate approximation of the lower-order gluon and ghost Green's functions Refs. \cite{Arbuzov:1980rm,Arbuzov:1982sz,Arbuzov:1987be}.
The expression in \eqn{gl_prop-guv} is used for the perturbative region, which is analogous to the \eqn{gl_prop2} in Ref. \cite{Davydychev:1996pb}. 
\\
In \fig{Lsymlat}, we plot our results using  Landau, Feynmann, Yennie and Arbuzov gauges for $L_{sym}(Q)$ in \eqn{Lsym-eq}.
We also compare with the results of lattice  taken from Refs.~\cite{Aguilar:2021lke, Athenodorou:2016oyh}  and a nonperturbative approximation of BC~\cite{Aguilar:2021lke}.
We predict that $L_{sym}(Q)$ demonstrates a zero in the Arbuzov gauge at $Q=0.034$ GeV, aligning with the critical value identified for the coupling constant $\alpha$ in \cite{Gracey:2023sup}. This value has previously been recognized for its particular importance in scenarios predominantly linked to QCD phenomenology or infrared dynamics.\\
\begin{figure}[htb]
\vspace{-0cm}
       \centerline{
       \includegraphics[scale=0.27,angle=0]{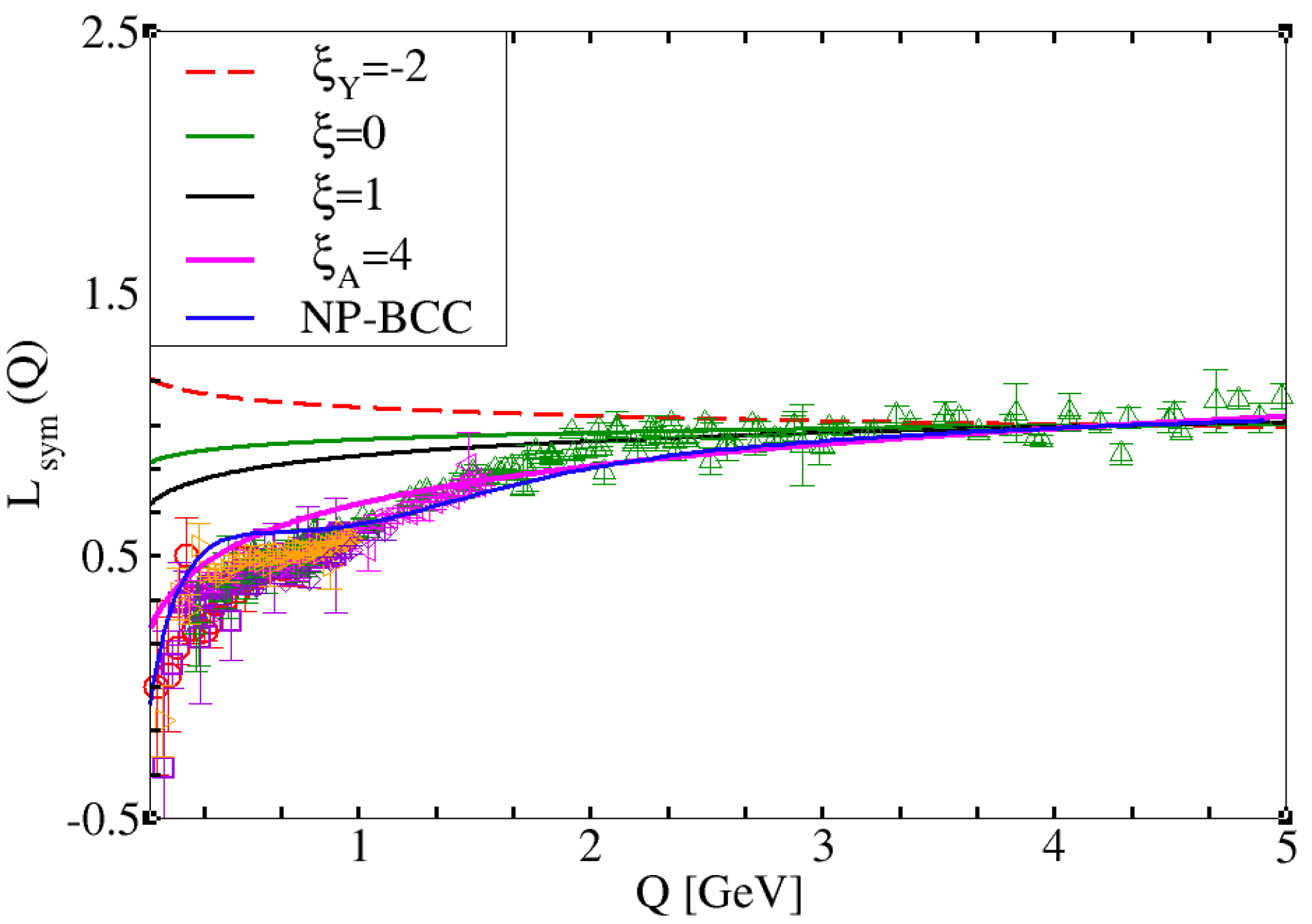}
       }
          \vspace{0cm}
       \caption{\label{Lsymlat} $L_{sym}(Q)$ in \eqn{Lsym-eq} compared with lattice, taken from Refs.~\cite{Aguilar:2021lke, Athenodorou:2016oyh} and other approach using a Nonperturbative BC construction (NP-BCC) of the three-gluon vertex~\cite{Aguilar:2019jsj}.  $L_{sym}(Q)$ exhibits a zero in the Arbuzov gauge at $Q=0.034$ GeV. }
\end{figure}
In \fig{Lsymyb}, we present a direct comparison between our results and the fit shown in \cite{Aguilar:2019jsj}; it is clear that the Arbuzov gauge is the one that best matches the results in the infrared, while the Yennie gauge should reproduce good results in the ultraviolet regimen.\\
Although calculated in the perturbative regime and extrapolated for small momenta, our results still exhibit the non-perturbative property on the zero crossing of the three-gluon vertex. In the Arbuzov gauge, the vertex has a zero for Q=0.034 GeV.
\begin{figure}[htb]
\vspace{-0cm}
       \centerline{
       \includegraphics[scale=0.27,angle=0]{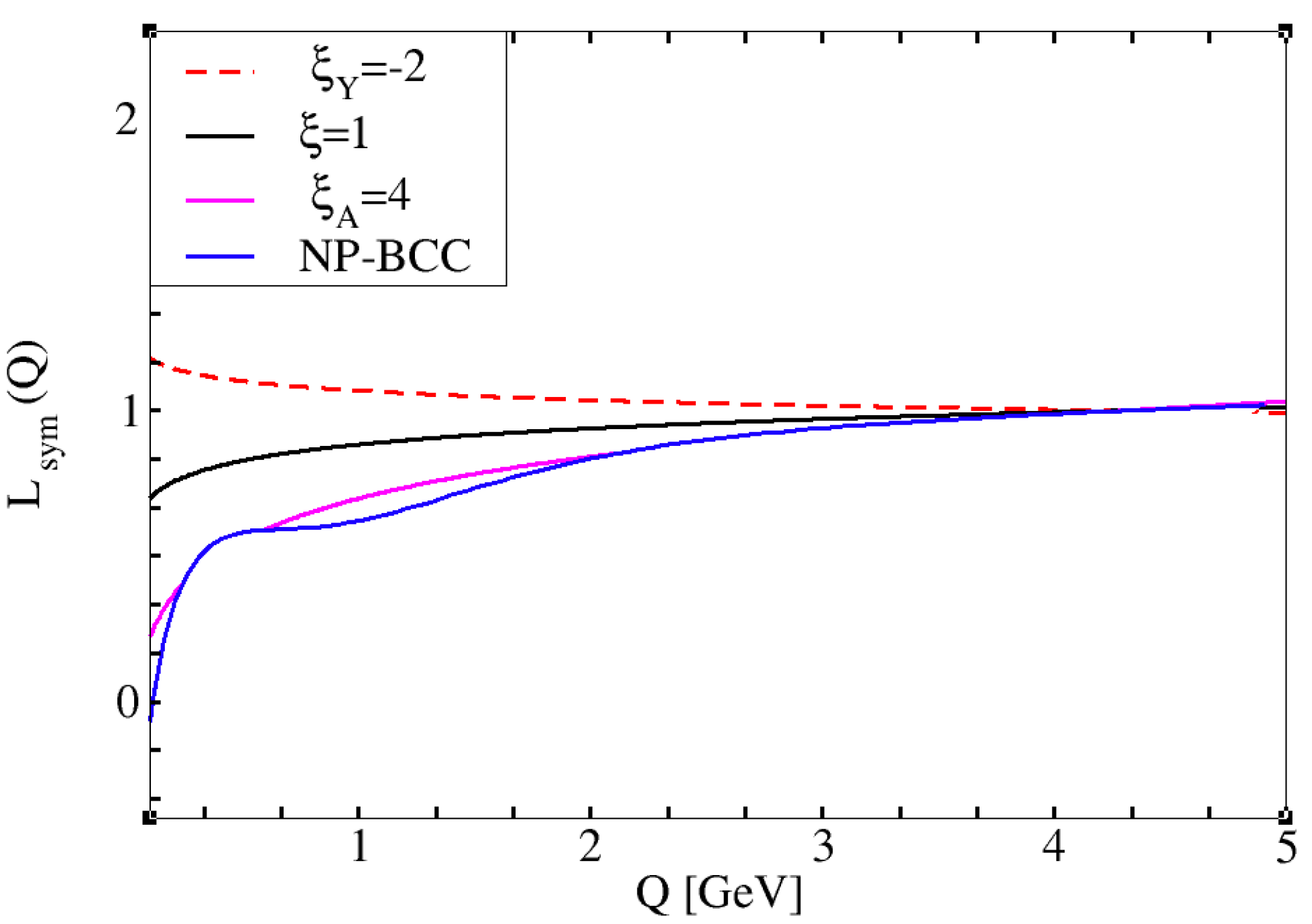}
       }
          \vspace{0cm}
       \caption{\label{Lsymyb} $L_{sym}(Q)$ in Landau, Yennie and Arbuzov  gauges at one-loop compared with nonperturbative results. The black line corresponds to the results reported in~\cite{Aguilar:2019jsj}.}
\end{figure}
Finally, in \fig{pd}, we show the  SDE result obtained in~\cite{Aguilar:2023qqd} and results of lattice three-gluon vertex in planar degeneracy~\cite{Pinto-Gomez:2022brg} in the range $[0, 5]$ GeV. In planar degeneracy~\cite{Aguilar:2023qqd,Pinto-Gomez:2022brg} there is only one kinematic variable $s^2=1/2(p^2+k^2+q^2)$ that practically determines the associated form factors, observing that at the symmetric limit $s^2=3Q^2/2$, given this outcome, our findings can be compared to those reported by ~\cite{Aguilar:2019jsj} in the manner described below,
\bea
\bar{\Gamma}_{1}^{\rm sym}(Q^2)\thickapprox L_{\rm sym}\left(\frac{3Q^2}{2}\right)\;.
\eea
\begin{figure}[htb]
\vspace{-0cm}
       \centerline{
       \includegraphics[scale=0.27,angle=0]{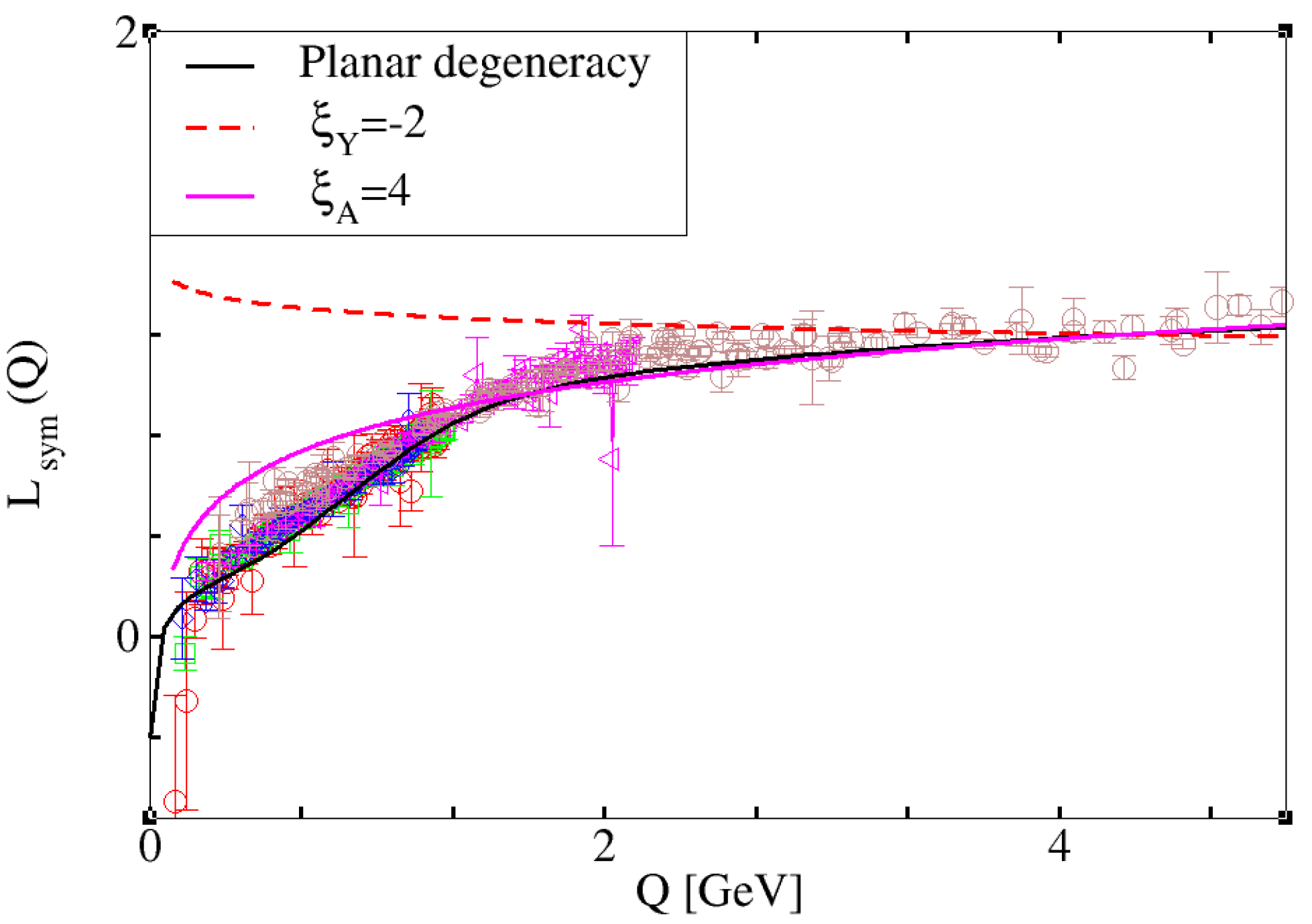}
       }
          \vspace{0cm}
       \caption{\label{pd} $L_{\rm sym}(Q)$ in Yennie and Arbuzov gauges at one-loop compared with the SDE result combined with planar degeneracy results in~\cite{Aguilar:2023qqd}. The symbols correspond to results in lattice~\cite{Pinto-Gomez:2022brg}, data are also displayed in terms of $Q^2$, for the sake of comparison.}
\end{figure}
Since the Yennie gauge cancels all ghost contributions at one-loop, it proves ineffective in this regime \cite{Davydychev:1996pb}. \fig{pd} indicates the infrared results obtained from lattice calculations do not align with those predicted by this gauge. Although the Yennie gauge should theoretically converge with lattice results in the ultraviolet region, current data do not yet permit meaningful comparisons at momenta exceeding 5 GeV.

Our analysis reveals that the planar degeneration is very close to the perturbative limit in Arbuzov gauge and in the symmetric momentum configuration, especially for momenta between [2,5] GeV, which is in complete agreement with the results reported in Ref.\cite{Aguilar:2023qqd}. The results in the Arbuzov and Yennie framework are valid only in the perturbative regime, although we have extrapolated them to very small momenta for comparison. Our findings show that our results are comparable to those of lattice simulations, but in our case, the calculations are reduced to only three form factors. In fact, our calculations demonstrate that one of these form factors dominates the entire contribution to the vertex. A considerable advantage of our work is the simplification of the equations, which allows us to extrapolate our results to both large and small momenta. The comprehensive analysis of simulations for the triple-gluon vertex using lattice methods is highly costly. As a consequence, the outcomes in the perturbative regime become crucial for validating the findings.
\section{Conclusions}
\label{Conclusions}
In the present work, we have conducted an analysis of the perturbative form factors of the three-gluon vertex at the one-loop level. For our study, we employed three different representations and provided the calculations for converting between them in the appendices. Our results are valid in the symmetric momenta configuration within the perturbative regime. Using the MOM renormalization scheme, only three of the fourteen structures that constitute the vertex (ten longitudinal and four transverse) are non-zero: $X_1$, $X_3$, and $Y_4$ in the AFFP representation, and $G_0$, $G_1$, and $G_2$ in the DOT representation. The structures for the three-gluon vertex are computed in our approximation for several gauges.\\
According to our findings, for positive gauges, the form factors of the three-gluon vertex increase in the infrared, while for negative gauges, they decrease. We identified a critical value $\xi_c = -0.89$ at which $L_{\text{sym}}(Q)$ remains almost constant in the infrared. 
From Figs. \ref{Lnormas}-\ref{pd}, it is evident that the Arbuzov gauge captures both the infrared and ultraviolet behaviors effectively, as the results align closely with those obtained from lattice QCD, including the presence of a zero-crossing. On the other hand, it is interesting to take a closer look at the Yennie gauge: although it does not match the lattice results in the infrared domain, instead showing a positive enhancement in this region and lacking a zero-crossing, it does provide reasonable results in the ultraviolet region as theoretically expected. This difference in behavior may indicate that the zero-crossing observed in lattice QCD results is influenced by ghost contributions, which are absent in the Yennie gauge at one-loop order due to their explicit cancellation.
However, a concise discussion of the origin of the zero-crossing would likely require a non-perturbative analysis, which is beyond the scope of this study.
Furthermore, these gauges simplify the calculations by reducing the divergences, primarily in the contribution of the ghosts.\\
The combination of the form factors presented here displays a zero crossing in the Arbuzov gauge at around 34 MeV. It is crucial to note that our computations maintain this characteristic qualitatively, even though they lack the power to produce non-perturbative results.\\
Lastly, we find that in this momenta configuration, our results are extremely close to those of the so-called planar degeneration. Given that the two configurations share the same properties, this cross-check is quite logical. In the near future, we plan to extend this work to other configurations of momenta, specifically asymmetric and orthogonal configurations.
\backmatter
\bmhead{Acknowledgements}
The author thanks A.C. Aguilar for providing the results of lattice and non-perturbative calculations.
L.X.G. wishes to thank National Council of Humanities, Sciences, and Technologies (CONAHCyT) for the support provided to her through the Cátedras CONAHCyT program and Project CBF2023-2024-268, Física Hadrónica en JLab: Descifrando la estructura interna de los mesones y bariones, from the 2023-2024 frontier science call.  A.R. acknowledges CIC-UMSNH for financial support under grant 18371.
The authors are also funded by sistema Nacional de Investigadoras e investigadores from CONAHCyT.
\begin{appendices}
\section{BC to DOT Representation}
\label{BC-D}
Comparison with the decomposition  \eqn{davy-b})  used in~\cite{Davydychev:1996pb}
gives the following representations of $Z$s \footnote{We have slightly changed  the notation used in~\cite{Davydychev:1996pb}, where $Z_{jkl}$ are scalar functions depending on $p^2, k^2$ and
$q^2$.} in terms of 
the functions in eqs. (\ref{BC-ggg-L}, \ref{BC-ggg-T}) used by Ball and Chiu \cite{Ball:1980ax}:
\bea
\nn Z_{1} &=& A(p^2, k^2; q^2) - (p k) C(p^2, k^2; q^2)
+ B(p^2, k^2; q^2)
         + (p k)(k q) F(p^2, k^2; q^2) - (k q) H ,
         \\
\nn Z_{2} &=& - 2 A(q^2, p^2; k^2) + 2 (p q) C(q^2, p^2; k^2)
+ k^2 (p q) F(q^2, p^2; k^2) - k^2 H ,
         \\
\nn Z_{3} &=& A(k^2, q^2; p^2) - (k q) C(k^2, q^2; p^2)
- B(k^2, q^2; p^2) + (p k)(k q) F(k^2, q^2; p^2) - (p k) H ,
\\
\nn Z_{4} &=& - A(p^2, k^2; q^2) \!+ (p k) C(p^2, k^2; q^2)
+ B(p^2, k^2; q^2) \!- (p k)(p q) F(p^2, k^2; q^2) \!+(p q) H ,
\\
\nn Z_{5} &=& - A(q^2, p^2; k^2) \!+ (p q) C(q^2, p^2; k^2)
- B(q^2, p^2; k^2)
         - (p k)(p q) F(q^2, p^2; k^2) \!+ (p k) H ,
\\
\nn Z_{6} &=& 2 A(k^2, q^2; p^2) - 2 (k q) C(k^2, q^2; p^2)
- p^2 (k q) F(k^2, q^2; p^2) + p^2 H ,
\\         
\nn Z_{7} &=& 2 C(q^2, p^2; k^2) + k^2 F(q^2, p^2; k^2),
\\
\nn Z_{8} &=& -2 C(k^2, q^2; p^2) - p^2 F(k^2, q^2; p^2),
\\
\nn Z_{9} &=& - C(k^2, q^2; p^2) + (p k) F(k^2, q^2; p^2)
          + H - S ,
\\
\nn Z_{10} &=& C(q^2, p^2; k^2) - (p k) F(q^2, p^2; k^2) ,
\\
\nn Z_{11} &=& C(p^2, k^2; q^2) + 2 C(q^2, p^2; k^2)
          - (k q) F(p^2, k^2; q^2)
          + k^2 F(q^2, p^2; k^2) - H - S ,
\\
\nn Z_{12} &=& - C(k^2, q^2; p^2) + (p k) F(k^2, q^2; p^2) ,
\\
\nn Z_{13} &=& -  C(p^2, k^2; q^2) - 2 C(k^2, q^2; p^2)
          + (p q) F(p^2, k^2; q^2)
          - p^2 F(k^2, q^2; p^2) + H - S ,
\\ \nn
 Z_{14} &=& C(q^2, p^2; k^2) - (p k) F(q^2, p^2; k^2)
          - H - S ,
          \eea
It is certain that renormalization is required for all structures that are dependent on function $A$. Notwithstanding the fact that we only display the symmetric limit in our computations, we provide the change of representation for a general momenta configuration.
\section{DOT to BC Representation}
\label{D-BC}
In this Appendix we briefly describe the expressions needed to change the representation from BC to DOT.
In terms of $Z$'s:
\bea \nn
S(p^2, k^2, q^2) &=&
\frac{1}{2}
\left\{ - Z_{9} + Z_{10} + Z_{12} - Z_{14} \right\} ,
\\ \nn
H(p^2, k^2, q^2) &=&
\frac{1}{2}
\left\{ Z_{9} + Z_{10} - Z_{12} - Z_{14} \right\} ,
\\ \nn
A(p^2, k^2; q^2) &=&
\frac{1}{2}\bigg\{ \frac{}{}
(p k) \left[ - Z_{7} + Z_{8} + Z_{11} - Z_{13} \right]
+ Z_{1} - Z_{4} - (p^2 + k^2) H  \bigg\} ,
\\ \nn
A(k^2, q^2; p^2) &=&
\frac{1}{2}
\left\{ - (k q) Z_{8} + Z_{6} - p^2 H \right\} ,
\\ \nn
A(q^2, p^2; k^2) &=&
\frac{1}{2}
\left\{ (p q) Z_{7} + Z_{2} - k^2 H \right\}\;,
\\
\nn B(p^2, k^2; q^2) &=&
\frac{1}{2} \bigg\{
p^2 \left[ Z_{9} - Z_{12} \right]
+ k^2 \left[ - Z_{10} + Z_{14} \right]
 - (p k) \left[ Z_{7} + Z_{8} - Z_{11} - Z_{13} \right]
+ Z_{1} + Z_{4} + q^2 S \bigg\} ,
\\ \nn
B(k^2, q^2; p^2) &=&
\frac{1}{2} 
\bigg\{-(k q)(Z_{8}-2 Z_{12}) + Z_{6} 
- 2 Z_{3}+ (k^2 - q^2) H \bigg\} ,
\\ 
\nn B(q^2, p^2; k^2) &=&
\frac{1}{2} 
\bigg\{-(p q)(Z_{7}-2 Z_{10}) + Z_{2} 
- 2 Z_{5}+ (q^2 - p^2) H \bigg\} ,
\\ \nn
C(p^2, k^2; q^2) &=&
\frac{1}{p^2 - k^2}
\bigg\{ (p q) \left[ Z_{7} - Z_{11} - Z_{10} + Z_{14} \right]
 + (k q) \left[ Z_{8} + Z_{9} - Z_{12} - Z_{13} \right]
\bigg\} ,
\\ \nn
C(k^2, q^2; p^2) &=&
\frac{1}{k^2 - q^2}
\left\{ (p k) Z_{8} + p^2 Z_{12} \right\}  ,
\\ \nn
C(q^2, p^2; k^2) &=&
\frac{1}{q^2 - p^2}
\left\{ (p k) Z_{7} + k^2 Z_{10} \right\}  ,
\\ \nn
F(p^2, k^2; q^2) &=&
\frac{1}{p^2 - k^2}
\bigg\{ Z_{7} + Z_{8} + Z_{9} - Z_{10} - Z_{11}
- Z_{12} - Z_{13} + Z_{14} \bigg\} ,
\\ \nn
F(k^2, q^2; p^2) &=&
\frac{1}{k^2 - q^2}
\left\{ Z_{8} - 2 Z_{12} \right\}  ,
\\ \
\nn F(q^2, p^2; k^2) &=&
\frac{1}{q^2 - p^2}
\left\{ Z_{7} - 2 Z_{10} \right\}  .
\eea
\\
\section{DOT to AFFP Basis}
\label{D-A}
We present the connection between the representation given in \cite{Aguilar:2019jsj} and that presented in \cite{Davydychev:1996pb}. The notation is in accordance with Section \ref{basis} of this paper
 \bea
\nn X_1&=& \frac{1}{4} \bigg[-((pq)+(pk)) (Z_{10}-Z_{12}
-Z_{14}+Z_{9})+2 (qk) Z_{8}-2 Z_{6}\bigg]
\\
\nn X_2&=&\frac{1}{4} (((pq)-(pk)) (Z_{10}-Z_{12}
- Z_{14}+Z_{9})+2 (qk) (2 Z_{12}-Z_{8})-4 Z_{3}+2 Z_{6})
\\
\nn X_3 &=&\frac{(pq) Z_{12}+(pk) (Z_{12}-Z_{8})}{(pq)-(pk)}
\\
\nn
X_4&=& \frac{1}{4} (2 ((pk) (-Z_{10}-Z_{11}+Z_{12}+Z_{13}+Z_{14}+Z_{7}-Z_{8}-Z_{9})-Z_{1}+Z_{4})+ q^2 (Z_{10}-Z_{12}-Z_{14}+Z_{9}))
\eea
\bea
\nn
X_5&=&\frac{1}{4}  (2 (pk) (Z_{10}+Z_{11}+Z_{12}+Z_{13}
-Z_{14}-Z_{7}-Z_{8}-Z_{9})+(q^2+2 (qk))\bigg[Z_{10}\\
\nn &-& Z_{12}-Z_{14}+Z_{9}\bigg]+2 (Z_{1}+Z_{4}))
\\
\nn
X_6&=&\frac{1}{(q^2+2 (qk))}q^2 (Z_{10}+Z_{11}-Z_{14}-Z_{7})+(qk) (Z_{10}+Z_{11}-Z_{12}-Z_{13}-Z_{14}-Z_{7}+Z_{8}+Z_{9}))
\\
\nn
X_7&=&\frac{1}{4}\bigg[-2 (pq) Z_{7}+k^2 \bigg(Z_{10}-Z_{12}
- Z_{14}+Z_{9}\bigg)+2 Z_{2}\bigg]
\eea
\bea
\nn X_8&=&\frac{1}{4} \bigg[(pq) (4 Z_{10}-2 Z_{7})-(2 (qk)+k^2) (Z_{10}-Z_{12}
- Z_{14}+Z_{9})+2 (Z_{2}-2 Z_{5})\bigg]
\\
\nn
X_9&=&\frac{(k^2 Z_{10}-Z_{7} ((qk)+k^2))}{(2 (qk)+k^2)}
\\
\nn
X_{10}&=&\frac{1}{2} (Z_{10}+Z_{12}-Z_{14}-Z_{9})
\\
\nn Y_1&=&\frac{ (Z_{8}-2 Z_{12})}{((pk)-(pq))}
\\
\nn
Y_2&=& \frac{(Z_{10}+Z_{11}+Z_{12}+Z_{13}-Z_{14}-Z_{7}-Z_{8}-Z_{9})}{(q^2+2 (qk))}
\\
\nn
Y_3&=&\frac{(Z_{7}-2 Z_{10})}{(2 (qk)+k^2)}
\\
\nn
Y_4&=& \frac{1}{2}(-Z_{10}+Z_{12}+Z_{14}-Z_{9})
\eea
\end{appendices}
\bibliography{ccc-a}


\begin{thebibliography}{55}
\ifx \bisbn   \undefined \def \bisbn  #1{ISBN #1}\fi
\ifx \binits  \undefined \def \binits#1{#1}\fi
\ifx \bauthor  \undefined \def \bauthor#1{#1}\fi
\ifx \batitle  \undefined \def \batitle#1{#1}\fi
\ifx \bjtitle  \undefined \def \bjtitle#1{#1}\fi
\ifx \bvolume  \undefined \def \bvolume#1{\textbf{#1}}\fi
\ifx \byear  \undefined \def \byear#1{#1}\fi
\ifx \bissue  \undefined \def \bissue#1{#1}\fi
\ifx \bfpage  \undefined \def \bfpage#1{#1}\fi
\ifx \blpage  \undefined \def \blpage #1{#1}\fi
\ifx \burl  \undefined \def \burl#1{\textsf{#1}}\fi
\ifx \doiurl  \undefined \def \doiurl#1{\url{https://doi.org/#1}}\fi
\ifx \betal  \undefined \def \betal{\textit{et al.}}\fi
\ifx \binstitute  \undefined \def \binstitute#1{#1}\fi
\ifx \binstitutionaled  \undefined \def \binstitutionaled#1{#1}\fi
\ifx \bctitle  \undefined \def \bctitle#1{#1}\fi
\ifx \beditor  \undefined \def \beditor#1{#1}\fi
\ifx \bpublisher  \undefined \def \bpublisher#1{#1}\fi
\ifx \bbtitle  \undefined \def \bbtitle#1{#1}\fi
\ifx \bedition  \undefined \def \bedition#1{#1}\fi
\ifx \bseriesno  \undefined \def \bseriesno#1{#1}\fi
\ifx \blocation  \undefined \def \blocation#1{#1}\fi
\ifx \bsertitle  \undefined \def \bsertitle#1{#1}\fi
\ifx \bsnm \undefined \def \bsnm#1{#1}\fi
\ifx \bsuffix \undefined \def \bsuffix#1{#1}\fi
\ifx \bparticle \undefined \def \bparticle#1{#1}\fi
\ifx \barticle \undefined \def \barticle#1{#1}\fi
\bibcommenthead
\ifx \bconfdate \undefined \def \bconfdate #1{#1}\fi
\ifx \botherref \undefined \def \botherref #1{#1}\fi
\ifx \url \undefined \def \url#1{\textsf{#1}}\fi
\ifx \bchapter \undefined \def \bchapter#1{#1}\fi
\ifx \bbook \undefined \def \bbook#1{#1}\fi
\ifx \bcomment \undefined \def \bcomment#1{#1}\fi
\ifx \oauthor \undefined \def \oauthor#1{#1}\fi
\ifx \citeauthoryear \undefined \def \citeauthoryear#1{#1}\fi
\ifx \endbibitem  \undefined \def \endbibitem {}\fi
\ifx \bconflocation  \undefined \def \bconflocation#1{#1}\fi
\ifx \arxivurl  \undefined \def \arxivurl#1{\textsf{#1}}\fi
\csname PreBibitemsHook\endcsname

\bibitem[\protect\citeauthoryear{Gross and Wilczek}{1973}]{PhysRevLett.30.1343}
\begin{barticle}
\bauthor{\bsnm{Gross}, \binits{D.J.}},
\bauthor{\bsnm{Wilczek}, \binits{F.}}:
\batitle{Ultraviolet behavior of non-abelian gauge theories}.
\bjtitle{Phys. Rev. Lett.}
\bvolume{30},
\bfpage{1343}--\blpage{1346}
(\byear{1973})
\doiurl{10.1103/PhysRevLett.30.1343}
\end{barticle}
\endbibitem

\bibitem[\protect\citeauthoryear{Politzer}{1973}]{PhysRevLett.30.1346}
\begin{barticle}
\bauthor{\bsnm{Politzer}, \binits{H.D.}}:
\batitle{Reliable perturbative results for strong interactions?}
\bjtitle{Phys. Rev. Lett.}
\bvolume{30},
\bfpage{1346}--\blpage{1349}
(\byear{1973})
\doiurl{10.1103/PhysRevLett.30.1346}
\end{barticle}
\endbibitem

\bibitem[\protect\citeauthoryear{Barber et~al.}{1979}]{Barber:1979yr}
\begin{barticle}
\bauthor{\bsnm{Barber}, \binits{D.P.}}, \betal:
\batitle{{Discovery of Three Jet Events and a Test of Quantum Chromodynamics at
  PETRA Energies}}.
\bjtitle{Phys. Rev. Lett.}
\bvolume{43},
\bfpage{830}
(\byear{1979})
\doiurl{10.1103/PhysRevLett.43.830}
\end{barticle}
\endbibitem

\bibitem[\protect\citeauthoryear{Brandelik et~al.}{1979}]{TASSO:1979zyf}
\begin{barticle}
\bauthor{\bsnm{Brandelik}, \binits{R.}}, \betal:
\batitle{{Evidence for Planar Events in e+ e- Annihilation at High-Energies}}.
\bjtitle{Phys. Lett. B}
\bvolume{86},
\bfpage{243}--\blpage{249}
(\byear{1979})
\doiurl{10.1016/0370-2693(79)90830-X}
\end{barticle}
\endbibitem

\bibitem[\protect\citeauthoryear{Ellis et~al.}{1976}]{Ellis:1976uc}
\begin{barticle}
\bauthor{\bsnm{Ellis}, \binits{J.R.}},
\bauthor{\bsnm{Gaillard}, \binits{M.K.}},
\bauthor{\bsnm{Ross}, \binits{G.G.}}:
\batitle{{Search for Gluons in e+ e- Annihilation}}.
\bjtitle{Nucl. Phys. B}
\bvolume{111},
\bfpage{253}
(\byear{1976})
\doiurl{10.1016/0550-3213(77)90253-X} .
\bcomment{[Erratum: Nucl.Phys.B 130, 516 (1977)]}
\end{barticle}
\endbibitem

\bibitem[\protect\citeauthoryear{Davydychev et~al.}{2000}]{PhysRevD.63.014022}
\begin{barticle}
\bauthor{\bsnm{Davydychev}, \binits{A.I.}},
\bauthor{\bsnm{Osland}, \binits{P.}},
\bauthor{\bsnm{Saks}, \binits{L.}}:
\batitle{Quark-gluon vertex in arbitrary gauge and dimension}.
\bjtitle{Phys. Rev. D}
\bvolume{63},
\bfpage{014022}
(\byear{2000})
\doiurl{10.1103/PhysRevD.63.014022}
\end{barticle}
\endbibitem

\bibitem[\protect\citeauthoryear{Skullerud et~al.}{2001}]{Skullerud:2001aw}
\begin{barticle}
\bauthor{\bsnm{Skullerud}, \binits{J.}},
\bauthor{\bsnm{Leinweber}, \binits{D.B.}},
\bauthor{\bsnm{Williams}, \binits{A.G.}}:
\batitle{{Nonperturbative improvement and tree level correction of the quark
  propagator}}.
\bjtitle{Phys. Rev. D}
\bvolume{64},
\bfpage{074508}
(\byear{2001})
\doiurl{10.1103/PhysRevD.64.074508}
{\href{https://arxiv.org/abs/hep-lat/0102013}{{arXiv:hep-lat/0102013}}}
\end{barticle}
\endbibitem

\bibitem[\protect\citeauthoryear{Kizilersu et~al.}{2007}]{Kizilersu:2006et}
\begin{barticle}
\bauthor{\bsnm{Kizilersu}, \binits{A.}},
\bauthor{\bsnm{Leinweber}, \binits{D.B.}},
\bauthor{\bsnm{Skullerud}, \binits{J.-I.}},
\bauthor{\bsnm{Williams}, \binits{A.G.}}:
\batitle{{Quark-gluon vertex in general kinematics}}.
\bjtitle{Eur. Phys. J. C}
\bvolume{50},
\bfpage{871}--\blpage{875}
(\byear{2007})
\doiurl{10.1140/epjc/s10052-007-0250-6}
{\href{https://arxiv.org/abs/hep-lat/0610078}{{arXiv:hep-lat/0610078}}}
\end{barticle}
\endbibitem

\bibitem[\protect\citeauthoryear{Oliveira et~al.}{2016}]{Oliveira:2016muq}
\begin{barticle}
\bauthor{\bsnm{Oliveira}, \binits{O.}},
\bauthor{\bsnm{K\i{}z\i{}lersu}, \binits{A.}},
\bauthor{\bsnm{Silva}, \binits{P.J.}},
\bauthor{\bsnm{Skullerud}, \binits{J.-I.}},
\bauthor{\bsnm{Sternbeck}, \binits{A.}},
\bauthor{\bsnm{Williams}, \binits{A.G.}}:
\batitle{{Lattice Landau gauge quark propagator and the quark-gluon vertex}}.
\bjtitle{Acta Phys. Polon. Supp.}
\bvolume{9},
\bfpage{363}--\blpage{368}
(\byear{2016})
\doiurl{10.5506/APhysPolBSupp.9.363}
{\href{https://arxiv.org/abs/1605.09632}{{arXiv:1605.09632}}}
{[hep-lat]}
\end{barticle}
\endbibitem

\bibitem[\protect\citeauthoryear{K\i{}z\i{}lers\"u
  et~al.}{2021}]{Kizilersu:2021jen}
\begin{barticle}
\bauthor{\bsnm{K\i{}z\i{}lers\"u}, \binits{A.}},
\bauthor{\bsnm{Oliveira}, \binits{O.}},
\bauthor{\bsnm{Silva}, \binits{P.J.}},
\bauthor{\bsnm{Skullerud}, \binits{J.-I.}},
\bauthor{\bsnm{Sternbeck}, \binits{A.}}:
\batitle{{Quark-gluon vertex from Nf=2 lattice QCD}}.
\bjtitle{Phys. Rev. D}
\bvolume{103}(\bissue{11}),
\bfpage{114515}
(\byear{2021})
\doiurl{10.1103/PhysRevD.103.114515}
{\href{https://arxiv.org/abs/2103.02945}{{arXiv:2103.02945}}}
{[hep-lat]}
\end{barticle}
\endbibitem

\bibitem[\protect\citeauthoryear{Oliveira et~al.}{2018}]{Oliveira:2018fkj}
\begin{barticle}
\bauthor{\bsnm{Oliveira}, \binits{O.}},
\bauthor{\bsnm{Frederico}, \binits{T.}},
\bauthor{\bsnm{Paula}, \binits{W.}},
\bauthor{\bsnm{Melo}, \binits{J.P.B.C.}}:
\batitle{{Exploring the Quark-Gluon Vertex with Slavnov-Taylor Identities and
  Lattice Simulations}}.
\bjtitle{Eur. Phys. J. C}
\bvolume{78}(\bissue{7}),
\bfpage{553}
(\byear{2018})
\doiurl{10.1140/epjc/s10052-018-6037-0}
{\href{https://arxiv.org/abs/1807.00675}{{arXiv:1807.00675}}}
{[hep-ph]}
\end{barticle}
\endbibitem

\bibitem[\protect\citeauthoryear{Davydychev et~al.}{1996}]{Davydychev:1996pb}
\begin{barticle}
\bauthor{\bsnm{Davydychev}, \binits{A.I.}},
\bauthor{\bsnm{Osland}, \binits{P.}},
\bauthor{\bsnm{Tarasov}, \binits{O.V.}}:
\batitle{{Three gluon vertex in arbitrary gauge and dimension}}.
\bjtitle{Phys. Rev. D}
\bvolume{54},
\bfpage{4087}--\blpage{4113}
(\byear{1996})
\doiurl{10.1103/PhysRevD.59.109901}
{\href{https://arxiv.org/abs/hep-ph/9605348}{{arXiv:hep-ph/9605348}}}.
\bcomment{[Erratum: Phys.Rev.D 59, 109901 (1999)]}
\end{barticle}
\endbibitem

\bibitem[\protect\citeauthoryear{Aguilar et~al.}{2023}]{Aguilar:2023mam}
\begin{barticle}
\bauthor{\bsnm{Aguilar}, \binits{A.C.}},
\bauthor{\bsnm{Ferreira}, \binits{M..N.}},
\bauthor{\bsnm{Iba\~nez}, \binits{D.}},
\bauthor{\bsnm{Papavassiliou}, \binits{J.}}:
\batitle{{Schwinger displacement of the quark\textendash{}gluon vertex}}.
\bjtitle{Eur. Phys. J. C}
\bvolume{83}(\bissue{10}),
\bfpage{967}
(\byear{2023})
\doiurl{10.1140/epjc/s10052-023-12103-8}
{\href{https://arxiv.org/abs/2308.16297}{{arXiv:2308.16297}}}
{[hep-ph]}
\end{barticle}
\endbibitem

\bibitem[\protect\citeauthoryear{Sultan et~al.}{2021}]{Sultan:2018tet}
\begin{barticle}
\bauthor{\bsnm{Sultan}, \binits{M.A.}},
\bauthor{\bsnm{Raya}, \binits{K.}},
\bauthor{\bsnm{Akram}, \binits{F.}},
\bauthor{\bsnm{Bashir}, \binits{A.}},
\bauthor{\bsnm{Masud}, \binits{B.}}:
\batitle{{Effect of the quark-gluon vertex on dynamical chiral symmetry
  breaking}}.
\bjtitle{Phys. Rev. D}
\bvolume{103}(\bissue{5}),
\bfpage{054036}
(\byear{2021})
\doiurl{10.1103/PhysRevD.103.054036}
{\href{https://arxiv.org/abs/1810.01396}{{arXiv:1810.01396}}}
{[nucl-th]}
\end{barticle}
\endbibitem

\bibitem[\protect\citeauthoryear{Bermudez et~al.}{2017}]{Bermudez:2017bpx}
\begin{barticle}
\bauthor{\bsnm{Bermudez}, \binits{R.}},
\bauthor{\bsnm{Albino}, \binits{L.}},
\bauthor{\bsnm{Guti\'errez-Guerrero}, \binits{L.X.}},
\bauthor{\bsnm{Tejeda-Yeomans}, \binits{M.E.}},
\bauthor{\bsnm{Bashir}, \binits{A.}}:
\batitle{{Quark-gluon Vertex: A Perturbation Theory Primer and Beyond}}.
\bjtitle{Phys. Rev. D}
\bvolume{95}(\bissue{3}),
\bfpage{034041}
(\byear{2017})
\doiurl{10.1103/PhysRevD.95.034041}
{\href{https://arxiv.org/abs/1702.04437}{{arXiv:1702.04437}}}
{[hep-ph]}
\end{barticle}
\endbibitem

\bibitem[\protect\citeauthoryear{Oliveira et~al.}{2019}]{Oliveira:2018ukh}
\begin{barticle}
\bauthor{\bsnm{Oliveira}, \binits{O.}},
\bauthor{\bsnm{Paula}, \binits{W.}},
\bauthor{\bsnm{Frederico}, \binits{T.}},
\bauthor{\bsnm{Melo}, \binits{J.P.B.C.}}:
\batitle{{The Quark-Gluon Vertex and the QCD Infrared Dynamics}}.
\bjtitle{Eur. Phys. J. C}
\bvolume{79}(\bissue{2}),
\bfpage{116}
(\byear{2019})
\doiurl{10.1140/epjc/s10052-019-6617-7}
{\href{https://arxiv.org/abs/1807.10348}{{arXiv:1807.10348}}}
{[hep-ph]}
\end{barticle}
\endbibitem

\bibitem[\protect\citeauthoryear{Williams}{2015}]{Williams:2014iea}
\begin{barticle}
\bauthor{\bsnm{Williams}, \binits{R.}}:
\batitle{{The quark-gluon vertex in Landau gauge bound-state studies}}.
\bjtitle{Eur. Phys. J. A}
\bvolume{51}(\bissue{5}),
\bfpage{57}
(\byear{2015})
\doiurl{10.1140/epja/i2015-15057-4}
{\href{https://arxiv.org/abs/1404.2545}{{arXiv:1404.2545}}}
{[hep-ph]}
\end{barticle}
\endbibitem

\bibitem[\protect\citeauthoryear{Albino et~al.}{2019}]{Albino:2018ncl}
\begin{barticle}
\bauthor{\bsnm{Albino}, \binits{L.}},
\bauthor{\bsnm{Bashir}, \binits{A.}},
\bauthor{\bsnm{Guerrero}, \binits{L.X.G.}},
\bauthor{\bsnm{Bennich}, \binits{B.E.}},
\bauthor{\bsnm{Rojas}, \binits{E.}}:
\batitle{{Transverse Takahashi Identities and Their Implications for Gauge
  Independent Dynamical Chiral Symmetry Breaking}}.
\bjtitle{Physical Review D}
\bvolume{100}(\bissue{5}),
\bfpage{054028}
(\byear{2019})
\doiurl{10.1103/PhysRevD.100.054028}
{\href{https://arxiv.org/abs/1812.02280}{{arXiv:1812.02280}}}
{[nucl-th]}
\end{barticle}
\endbibitem

\bibitem[\protect\citeauthoryear{Albino et~al.}{2022}]{Albino:2022gzs}
\begin{barticle}
\bauthor{\bsnm{Albino}, \binits{L.}},
\bauthor{\bsnm{Higuera-Angulo}, \binits{I.M.}},
\bauthor{\bsnm{Raya}, \binits{K.}},
\bauthor{\bsnm{Bashir}, \binits{A.}}:
\batitle{{Pseudoscalar mesons: Light front wave functions, GPDs, and PDFs}}.
\bjtitle{Phys. Rev. D}
\bvolume{106}(\bissue{3}),
\bfpage{034003}
(\byear{2022})
\doiurl{10.1103/PhysRevD.106.034003}
{\href{https://arxiv.org/abs/2207.06550}{{arXiv:2207.06550}}}
{[hep-ph]}
\end{barticle}
\endbibitem

\bibitem[\protect\citeauthoryear{Binosi and
  Papavassiliou}{2018}]{Binosi:2017rwj}
\begin{barticle}
\bauthor{\bsnm{Binosi}, \binits{D.}},
\bauthor{\bsnm{Papavassiliou}, \binits{J.}}:
\batitle{{Coupled dynamics in gluon mass generation and the impact of the
  three-gluon vertex}}.
\bjtitle{Phys. Rev.}
\bvolume{D97}(\bissue{5}),
\bfpage{054029}
(\byear{2018})
\doiurl{10.1103/PhysRevD.97.054029}
{\href{https://arxiv.org/abs/1709.09964}{{arXiv:1709.09964}}}
{[hep-ph]}
\end{barticle}
\endbibitem

\bibitem[\protect\citeauthoryear{Blum et~al.}{2015}]{Blum:2015lsa}
\begin{barticle}
\bauthor{\bsnm{Blum}, \binits{A.L.}},
\bauthor{\bsnm{Alkofer}, \binits{R.}},
\bauthor{\bsnm{Huber}, \binits{M.Q.}},
\bauthor{\bsnm{Windisch}, \binits{A.}}:
\batitle{{Unquenching the three-gluon vertex: A status report}}.
\bjtitle{Acta Phys. Polon. Supp.}
\bvolume{8}(\bissue{2}),
\bfpage{321}
(\byear{2015})
\doiurl{10.5506/APhysPolBSupp.8.321}
{\href{https://arxiv.org/abs/1506.04275}{{arXiv:1506.04275}}}
{[hep-ph]}
\end{barticle}
\endbibitem

\bibitem[\protect\citeauthoryear{Eichmann et~al.}{2014}]{Eichmann:2014xya}
\begin{barticle}
\bauthor{\bsnm{Eichmann}, \binits{G.}},
\bauthor{\bsnm{Williams}, \binits{R.}},
\bauthor{\bsnm{Alkofer}, \binits{R.}},
\bauthor{\bsnm{Vujinovic}, \binits{M.}}:
\batitle{{Three-gluon vertex in Landau gauge}}.
\bjtitle{Phys. Rev. D}
\bvolume{89}(\bissue{10}),
\bfpage{105014}
(\byear{2014})
\doiurl{10.1103/PhysRevD.89.105014}
{\href{https://arxiv.org/abs/1402.1365}{{arXiv:1402.1365}}}
{[hep-ph]}
\end{barticle}
\endbibitem

\bibitem[\protect\citeauthoryear{Binger and Brodsky}{2006}]{Binger:2006sj}
\begin{barticle}
\bauthor{\bsnm{Binger}, \binits{M.}},
\bauthor{\bsnm{Brodsky}, \binits{S.J.}}:
\batitle{{The Form-factors of the gauge-invariant three-gluon vertex}}.
\bjtitle{Phys. Rev. D}
\bvolume{74},
\bfpage{054016}
(\byear{2006})
\doiurl{10.1103/PhysRevD.74.054016}
{\href{https://arxiv.org/abs/hep-ph/0602199}{{arXiv:hep-ph/0602199}}}
\end{barticle}
\endbibitem

\bibitem[\protect\citeauthoryear{Ball and Chiu}{1980}]{Ball:1980ax}
\begin{barticle}
\bauthor{\bsnm{Ball}, \binits{J.S.}},
\bauthor{\bsnm{Chiu}, \binits{T.-W.}}:
\batitle{{Analytic Properties of the Vertex Function in Gauge Theories. 2.}}
\bjtitle{Phys. Rev. D}
\bvolume{22},
\bfpage{2550}
(\byear{1980})
\doiurl{10.1103/PhysRevD.22.2550} .
\bcomment{[Erratum:Physical Review D 23, 3085 (1981)]}
\end{barticle}
\endbibitem

\bibitem[\protect\citeauthoryear{Aguilar et~al.}{2023}]{Aguilar:2023qqd}
\begin{barticle}
\bauthor{\bsnm{Aguilar}, \binits{A.C.}},
\bauthor{\bsnm{Ferreira}, \binits{M.N.}},
\bauthor{\bsnm{Papavassiliou}, \binits{J.}},
\bauthor{\bsnm{Santos}, \binits{L.R.}}:
\batitle{{Planar degeneracy of the three-gluon vertex}}.
\bjtitle{Eur. Phys. J. C}
\bvolume{83}(\bissue{6}),
\bfpage{549}
(\byear{2023})
\doiurl{10.1140/epjc/s10052-023-11732-3}
{\href{https://arxiv.org/abs/2305.05704}{{arXiv:2305.05704}}}
{[hep-ph]}
\end{barticle}
\endbibitem

\bibitem[\protect\citeauthoryear{Pinto-G\'omez
  et~al.}{2023}]{Pinto-Gomez:2022brg}
\begin{barticle}
\bauthor{\bsnm{Pinto-G\'omez}, \binits{F.}},
\bauthor{\bsnm{De~Soto}, \binits{F.}},
\bauthor{\bsnm{Ferreira}, \binits{M.N.}},
\bauthor{\bsnm{Papavassiliou}, \binits{J.}},
\bauthor{\bsnm{Rodr\'\i{}guez-Quintero}, \binits{J.}}:
\batitle{{Lattice three-gluon vertex in extended kinematics: Planar
  degeneracy}}.
\bjtitle{Phys. Lett. B}
\bvolume{838},
\bfpage{137737}
(\byear{2023})
\doiurl{10.1016/j.physletb.2023.137737}
{\href{https://arxiv.org/abs/2208.01020}{{arXiv:2208.01020}}}
{[hep-ph]}
\end{barticle}
\endbibitem

\bibitem[\protect\citeauthoryear{Skullerud et~al.}{2003}]{Skullerud:2003qu}
\begin{barticle}
\bauthor{\bsnm{Skullerud}, \binits{J.I.}},
\bauthor{\bsnm{Bowman}, \binits{P.O.}},
\bauthor{\bsnm{Kizilersu}, \binits{A.}},
\bauthor{\bsnm{Leinweber}, \binits{D.B.}},
\bauthor{\bsnm{Williams}, \binits{A.G.}}:
\batitle{{Nonperturbative structure of the quark gluon vertex}}.
\bjtitle{JHEP}
\bvolume{04},
\bfpage{047}
(\byear{2003})
\doiurl{10.1088/1126-6708/2003/04/047}
{\href{https://arxiv.org/abs/hep-ph/0303176}{{arXiv:hep-ph/0303176}}}
\end{barticle}
\endbibitem

\bibitem[\protect\citeauthoryear{Cucchieri et~al.}{2006}]{Cucchieri:2006tf}
\begin{barticle}
\bauthor{\bsnm{Cucchieri}, \binits{A.}},
\bauthor{\bsnm{Maas}, \binits{A.}},
\bauthor{\bsnm{Mendes}, \binits{T.}}:
\batitle{{Exploratory study of three-point Green's functions in Landau-gauge
  Yang-Mills theory}}.
\bjtitle{Phys. Rev. D}
\bvolume{74},
\bfpage{014503}
(\byear{2006})
\doiurl{10.1103/PhysRevD.74.014503}
{\href{https://arxiv.org/abs/hep-lat/0605011}{{arXiv:hep-lat/0605011}}}
\end{barticle}
\endbibitem

\bibitem[\protect\citeauthoryear{Athenodorou
  et~al.}{2016}]{Athenodorou:2016oyh}
\begin{barticle}
\bauthor{\bsnm{Athenodorou}, \binits{A.}},
\bauthor{\bsnm{Binosi}, \binits{D.}},
\bauthor{\bsnm{Boucaud}, \binits{P.}},
\bauthor{\bsnm{De~Soto}, \binits{F.}},
\bauthor{\bsnm{Papavassiliou}, \binits{J.}},
\bauthor{\bsnm{Rodriguez-Quintero}, \binits{J.}},
\bauthor{\bsnm{Zafeiropoulos}, \binits{S.}}:
\batitle{{On the zero crossing of the three-gluon vertex}}.
\bjtitle{Phys. Lett. B}
\bvolume{761},
\bfpage{444}--\blpage{449}
(\byear{2016})
\doiurl{10.1016/j.physletb.2016.08.065}
{\href{https://arxiv.org/abs/1607.01278}{{arXiv:1607.01278}}}
{[hep-ph]}
\end{barticle}
\endbibitem

\bibitem[\protect\citeauthoryear{Boucaud et~al.}{2017}]{Boucaud:2017obn}
\begin{barticle}
\bauthor{\bsnm{Boucaud}, \binits{P.}},
\bauthor{\bsnm{De~Soto}, \binits{F.}},
\bauthor{\bsnm{Rodr\'\i{}guez-Quintero}, \binits{J.}},
\bauthor{\bsnm{Zafeiropoulos}, \binits{S.}}:
\batitle{{Refining the detection of the zero crossing for the three-gluon
  vertex in symmetric and asymmetric momentum subtraction schemes}}.
\bjtitle{Phys. Rev. D}
\bvolume{95}(\bissue{11}),
\bfpage{114503}
(\byear{2017})
\doiurl{10.1103/PhysRevD.95.114503}
{\href{https://arxiv.org/abs/1701.07390}{{arXiv:1701.07390}}}
{[hep-lat]}
\end{barticle}
\endbibitem

\bibitem[\protect\citeauthoryear{Cucchieri et~al.}{2008}]{Cucchieri:2008qm}
\begin{barticle}
\bauthor{\bsnm{Cucchieri}, \binits{A.}},
\bauthor{\bsnm{Maas}, \binits{A.}},
\bauthor{\bsnm{Mendes}, \binits{T.}}:
\batitle{{Three-point vertices in Landau-gauge Yang-Mills theory}}.
\bjtitle{Phys. Rev. D}
\bvolume{77},
\bfpage{094510}
(\byear{2008})
\doiurl{10.1103/PhysRevD.77.094510}
{\href{https://arxiv.org/abs/0803.1798}{{arXiv:0803.1798}}}
{[hep-lat]}
\end{barticle}
\endbibitem

\bibitem[\protect\citeauthoryear{Duarte et~al.}{2016}]{Duarte:2016ieu}
\begin{barticle}
\bauthor{\bsnm{Duarte}, \binits{A.G.}},
\bauthor{\bsnm{Oliveira}, \binits{O.}},
\bauthor{\bsnm{Silva}, \binits{P.J.}}:
\batitle{{Further Evidence For Zero Crossing On The Three Gluon Vertex}}.
\bjtitle{Phys. Rev. D}
\bvolume{94}(\bissue{7}),
\bfpage{074502}
(\byear{2016})
\doiurl{10.1103/PhysRevD.94.074502}
{\href{https://arxiv.org/abs/1607.03831}{{arXiv:1607.03831}}}
{[hep-lat]}
\end{barticle}
\endbibitem

\bibitem[\protect\citeauthoryear{Fischer and Williams}{2009}]{Fischer:2009jm}
\begin{barticle}
\bauthor{\bsnm{Fischer}, \binits{C.S.}},
\bauthor{\bsnm{Williams}, \binits{R.}}:
\batitle{{Probing the gluon self-interaction in light mesons}}.
\bjtitle{Phys. Rev. Lett.}
\bvolume{103},
\bfpage{122001}
(\byear{2009})
\doiurl{10.1103/PhysRevLett.103.122001}
{\href{https://arxiv.org/abs/0905.2291}{{arXiv:0905.2291}}}
{[hep-ph]}
\end{barticle}
\endbibitem

\bibitem[\protect\citeauthoryear{Pelaez and Rios}{2006}]{Pelaez:2006nj}
\begin{barticle}
\bauthor{\bsnm{Pelaez}, \binits{J.R.}},
\bauthor{\bsnm{Rios}, \binits{G.}}:
\batitle{{Nature of the f0(600) from its N(c) dependence at two loops in
  unitarized Chiral Perturbation Theory}}.
\bjtitle{Phys. Rev. Lett.}
\bvolume{97},
\bfpage{242002}
(\byear{2006})
\doiurl{10.1103/PhysRevLett.97.242002}
{\href{https://arxiv.org/abs/hep-ph/0610397}{{arXiv:hep-ph/0610397}}}
\end{barticle}
\endbibitem

\bibitem[\protect\citeauthoryear{Aguilar et~al.}{2019}]{Aguilar:2019jsj}
\begin{barticle}
\bauthor{\bsnm{Aguilar}, \binits{A.C.}},
\bauthor{\bsnm{Ferreira}, \binits{M.N.}},
\bauthor{\bsnm{Figueiredo}, \binits{C.T.}},
\bauthor{\bsnm{Papavassiliou}, \binits{J.}}:
\batitle{{Nonperturbative Ball-Chiu construction of the three-gluon vertex}}.
\bjtitle{Phys. Rev. D}
\bvolume{99}(\bissue{9}),
\bfpage{094010}
(\byear{2019})
\doiurl{10.1103/PhysRevD.99.094010}
{\href{https://arxiv.org/abs/1903.01184}{{arXiv:1903.01184}}}
{[hep-ph]}
\end{barticle}
\endbibitem

\bibitem[\protect\citeauthoryear{Aguilar et~al.}{2021}]{Aguilar:2021lke}
\begin{barticle}
\bauthor{\bsnm{Aguilar}, \binits{A.C.}},
\bauthor{\bsnm{De~Soto}, \binits{F.}},
\bauthor{\bsnm{Ferreira}, \binits{M.N.}},
\bauthor{\bsnm{Papavassiliou}, \binits{J.}},
\bauthor{\bsnm{Rodr\'\i{}guez-Quintero}, \binits{J.}}:
\batitle{{Infrared facets of the three-gluon vertex}}.
\bjtitle{Phys. Lett. B}
\bvolume{818},
\bfpage{136352}
(\byear{2021})
\doiurl{10.1016/j.physletb.2021.136352}
{\href{https://arxiv.org/abs/2102.04959}{{arXiv:2102.04959}}}
{[hep-ph]}
\end{barticle}
\endbibitem

\bibitem[\protect\citeauthoryear{Papavassiliou
  et~al.}{2022}]{Papavassiliou:2022umz}
\begin{barticle}
\bauthor{\bsnm{Papavassiliou}, \binits{J.}},
\bauthor{\bsnm{Aguilar}, \binits{A.C.}},
\bauthor{\bsnm{Ferreira}, \binits{M.N.}}:
\batitle{{Theory and phenomenology of the three-gluon vertex}}.
\bjtitle{Rev. Mex. Fis. Suppl.}
\bvolume{3}(\bissue{3}),
\bfpage{0308112}
(\byear{2022})
\doiurl{10.31349/SuplRevMexFis.3.0308112}
{\href{https://arxiv.org/abs/2201.08496}{{arXiv:2201.08496}}}
{[hep-ph]}
\end{barticle}
\endbibitem

\bibitem[\protect\citeauthoryear{Parrinello}{1994}]{Parrinello:1994wd}
\begin{barticle}
\bauthor{\bsnm{Parrinello}, \binits{C.}}:
\batitle{{Exploratory study of the three gluon vertex on the lattice}}.
\bjtitle{Phys. Rev. D}
\bvolume{50},
\bfpage{4247}--\blpage{4251}
(\byear{1994})
\doiurl{10.1103/PhysRevD.50.R4247}
{\href{https://arxiv.org/abs/hep-lat/9405024}{{arXiv:hep-lat/9405024}}}
\end{barticle}
\endbibitem

\bibitem[\protect\citeauthoryear{Arbuzov et~al.}{1981}]{Arbuzov:1980rm}
\begin{barticle}
\bauthor{\bsnm{Arbuzov}, \binits{B.A.}},
\bauthor{\bsnm{Boos}, \binits{E.E.}},
\bauthor{\bsnm{Kurennoy}, \binits{S.S.}}:
\batitle{{Scale Solutions and Coupling Constant in Electrodynamics of Vector
  Particles}}.
\bjtitle{Yad. Fiz.}
\bvolume{34},
\bfpage{406}
(\byear{1981})
\end{barticle}
\endbibitem

\bibitem[\protect\citeauthoryear{Arbuzov}{1983}]{Arbuzov:1982sz}
\begin{barticle}
\bauthor{\bsnm{Arbuzov}, \binits{B.A.}}:
\batitle{{Infrared Asymptotics of Gluon and Quark Propagators in {QCD}}}.
\bjtitle{Phys. Lett. B}
\bvolume{125},
\bfpage{497}--\blpage{500}
(\byear{1983})
\doiurl{10.1016/0370-2693(83)91334-5}
\end{barticle}
\endbibitem

\bibitem[\protect\citeauthoryear{Arbuzov et~al.}{1988}]{Arbuzov:1986xu}
\begin{barticle}
\bauthor{\bsnm{Arbuzov}, \binits{B.A.}},
\bauthor{\bsnm{Boos}, \binits{E.E.}},
\bauthor{\bsnm{Davydychev}, \binits{A.I.}}:
\batitle{{Infrared Asymptotics of Gluon Green's Functions in Covariant Gauge}}.
\bjtitle{Theor. Math. Phys.}
\bvolume{74},
\bfpage{103}--\blpage{108}
(\byear{1988})
\doiurl{10.1007/BF01886478}
\end{barticle}
\endbibitem

\bibitem[\protect\citeauthoryear{Arbuzov}{1988}]{Arbuzov:1987be}
\begin{barticle}
\bauthor{\bsnm{Arbuzov}, \binits{B.A.}}:
\batitle{{QUANTUM CHROMODYNAMICS AT LARGE DISTANCES}}.
\bjtitle{Sov. J. Part. Nucl.}
\bvolume{19},
\bfpage{1}
(\byear{1988})
\end{barticle}
\endbibitem

\bibitem[\protect\citeauthoryear{Yennie et~al.}{1961}]{Yennie:1961ad}
\begin{barticle}
\bauthor{\bsnm{Yennie}, \binits{D.R.}},
\bauthor{\bsnm{Frautschi}, \binits{S.C.}},
\bauthor{\bsnm{Suura}, \binits{H.}}:
\batitle{{The infrared divergence phenomena and high-energy processes}}.
\bjtitle{Annals Phys.}
\bvolume{13},
\bfpage{379}--\blpage{452}
(\byear{1961})
\doiurl{10.1016/0003-4916(61)90151-8}
\end{barticle}
\endbibitem

\bibitem[\protect\citeauthoryear{Mizher et~al.}{2024}]{Mizher:2024zag}
\begin{barticle}
\bauthor{\bsnm{Mizher}, \binits{A.}},
\bauthor{\bsnm{Raya}, \binits{A.}},
\bauthor{\bsnm{Raya}, \binits{K.}}:
\batitle{{Fried-Yennie Gauge in Pseudo-QED}}.
\bjtitle{Entropy}
\bvolume{26}(\bissue{2}),
\bfpage{157}
(\byear{2024})
\doiurl{10.3390/e26020157}
{\href{https://arxiv.org/abs/2401.11964}{{arXiv:2401.11964}}}
{[hep-ph]}
\end{barticle}
\endbibitem

\bibitem[\protect\citeauthoryear{Gracey et~al.}{2023}]{Gracey:2023unc}
\begin{barticle}
\bauthor{\bsnm{Gracey}, \binits{J.A.}},
\bauthor{\bsnm{Mason}, \binits{R.H.}},
\bauthor{\bsnm{Ryttov}, \binits{T.A.}},
\bauthor{\bsnm{Simms}, \binits{R.M.}}:
\batitle{{Scheme and gauge dependence of QCD fixed points at five loops}}.
\bjtitle{Phys. Rev. D}
\bvolume{108}(\bissue{4}),
\bfpage{045006}
(\byear{2023})
\doiurl{10.1103/PhysRevD.108.045006}
{\href{https://arxiv.org/abs/2306.09056}{{arXiv:2306.09056}}}
{[hep-ph]}
\end{barticle}
\endbibitem

\bibitem[\protect\citeauthoryear{Gracey and Mason}{2023}]{Gracey:2023sup}
\begin{barticle}
\bauthor{\bsnm{Gracey}, \binits{J.A.}},
\bauthor{\bsnm{Mason}, \binits{R.H.}}:
\batitle{{Crewther\textquoteright{}s relation, schemes, gauges, and fixed
  points}}.
\bjtitle{Phys. Rev. D}
\bvolume{108}(\bissue{5}),
\bfpage{056006}
(\byear{2023})
\doiurl{10.1103/PhysRevD.108.056006}
{\href{https://arxiv.org/abs/2306.11416}{{arXiv:2306.11416}}}
{[hep-ph]}
\end{barticle}
\endbibitem

\bibitem[\protect\citeauthoryear{Boos and Davydychev}{1988}]{Boos1988}
\begin{botherref}
\oauthor{\bsnm{Boos}, \binits{E.E.}},
\oauthor{\bsnm{Davydychev}, \binits{A.I.}}:
Infrared problems of describing the glueball and an estimation of its mass.
INP MSU 88-021/42
(1988)
\end{botherref}
\endbibitem

\bibitem[\protect\citeauthoryear{Slavnov}{1972}]{Slavnov:1972fg}
\begin{barticle}
\bauthor{\bsnm{Slavnov}, \binits{A.A.}}:
\batitle{{Ward Identities in Gauge Theories}}.
\bjtitle{Theor. Math. Phys.}
\bvolume{10},
\bfpage{99}--\blpage{107}
(\byear{1972})
\doiurl{10.1007/BF01090719}
\end{barticle}
\endbibitem

\bibitem[\protect\citeauthoryear{Taylor}{1971}]{Taylor:1971ff}
\begin{barticle}
\bauthor{\bsnm{Taylor}, \binits{J.C.}}:
\batitle{{Ward Identities and Charge Renormalization of the Yang-Mills Field}}.
\bjtitle{Nucl. Phys. B}
\bvolume{33},
\bfpage{436}--\blpage{444}
(\byear{1971})
\doiurl{10.1016/0550-3213(71)90297-5}
\end{barticle}
\endbibitem

\bibitem[\protect\citeauthoryear{Aguilar et~al.}{2019}]{Aguilar:2018csq}
\begin{barticle}
\bauthor{\bsnm{Aguilar}, \binits{A.C.}},
\bauthor{\bsnm{Ferreira}, \binits{M.N.}},
\bauthor{\bsnm{Figueiredo}, \binits{C.T.}},
\bauthor{\bsnm{Papavassiliou}, \binits{J.}}:
\batitle{{Nonperturbative structure of the ghost-gluon kernel}}.
\bjtitle{Phys. Rev. D}
\bvolume{99}(\bissue{3}),
\bfpage{034026}
(\byear{2019})
\doiurl{10.1103/PhysRevD.99.034026}
{\href{https://arxiv.org/abs/1811.08961}{{arXiv:1811.08961}}}
{[hep-ph]}
\end{barticle}
\endbibitem

\bibitem[\protect\citeauthoryear{Celmaster and
  Gonsalves}{1979}]{Celmaster:1979km}
\begin{barticle}
\bauthor{\bsnm{Celmaster}, \binits{W.}},
\bauthor{\bsnm{Gonsalves}, \binits{R.J.}}:
\batitle{{The Renormalization Prescription Dependence of the QCD Coupling
  Constant}}.
\bjtitle{Phys. Rev. D}
\bvolume{20},
\bfpage{1420}
(\byear{1979})
\doiurl{10.1103/PhysRevD.20.1420}
\end{barticle}
\endbibitem

\bibitem[\protect\citeauthoryear{Grassi et~al.}{2001}]{Grassi:1999tp}
\begin{barticle}
\bauthor{\bsnm{Grassi}, \binits{P.A.}},
\bauthor{\bsnm{Hurth}, \binits{T.}},
\bauthor{\bsnm{Steinhauser}, \binits{M.}}:
\batitle{{Practical algebraic renormalization}}.
\bjtitle{Annals Phys.}
\bvolume{288},
\bfpage{197}--\blpage{248}
(\byear{2001})
\doiurl{10.1006/aphy.2001.6117}
{\href{https://arxiv.org/abs/hep-ph/9907426}{{arXiv:hep-ph/9907426}}}
\end{barticle}
\endbibitem

\bibitem[\protect\citeauthoryear{Fried and Yennie}{1958}]{Fried:1958zz}
\begin{barticle}
\bauthor{\bsnm{Fried}, \binits{H.M.}},
\bauthor{\bsnm{Yennie}, \binits{D.R.}}:
\batitle{{New Techniques in the Lamb Shift Calculation}}.
\bjtitle{Phys. Rev.}
\bvolume{112},
\bfpage{1391}--\blpage{1404}
(\byear{1958})
\doiurl{10.1103/PhysRev.112.1391}
\end{barticle}
\endbibitem

\bibitem[\protect\citeauthoryear{Sapirstein et~al.}{1984}]{Sapirstein:1983xr}
\begin{barticle}
\bauthor{\bsnm{Sapirstein}, \binits{J.R.}},
\bauthor{\bsnm{Terray}, \binits{E.A.}},
\bauthor{\bsnm{Yennie}, \binits{D.R.}}:
\batitle{{Radiative Recoil Corrections to Muonium and Positronium Hyperfine
  Splitting}}.
\bjtitle{Phys. Rev. D}
\bvolume{29},
\bfpage{2290}
(\byear{1984})
\doiurl{10.1103/PhysRevD.29.2290}
\end{barticle}
\endbibitem

\bibitem[\protect\citeauthoryear{Eides and Shelyuto}{2001}]{Eides:2001dw}
\begin{barticle}
\bauthor{\bsnm{Eides}, \binits{M.I.}},
\bauthor{\bsnm{Shelyuto}, \binits{V.A.}}:
\batitle{{One loop electron vertex in Yennie gauge}}.
\bjtitle{Eur. Phys. J. C}
\bvolume{21},
\bfpage{489}--\blpage{494}
(\byear{2001})
\doiurl{10.1007/s100520100745}
{\href{https://arxiv.org/abs/hep-ph/0102050}{{arXiv:hep-ph/0102050}}}
\end{barticle}
\endbibitem

\end{thebibliography}
\end{document}